\documentclass[fleqn,usenatbib]{mnras}

\usepackage[T1]{fontenc}
\usepackage{bm}
\usepackage{xcolor}
\DeclareRobustCommand{\VAN}[3]{#2}
\let\VANthebibliography\thebibliography
\def\thebibliography{\DeclareRobustCommand{\VAN}[3]{##3}\VANthebibliography}

\usepackage{graphicx}	
\usepackage{amsmath}	
\usepackage{cuted}
\usepackage{xcolor}
\usepackage{pdflscape}
\usepackage{tablefootnote}
\usepackage{threeparttable}

\title[DAmodel]{DAmodel: Hierarchical Bayesian Modelling of DA White Dwarfs for Spectrophotometric Calibration}

\author[B. M. Boyd et al.]{Benjamin M. Boyd$^{1}$\thanks{E-mail: \href{mailto:bmb41@cam.ac.uk}{bmb41@cam.ac.uk}}\orcidlink{0000-0002-0622-1117
}, Gautham Narayan$^{2}$\orcidlink{0000-0001-6022-0484}, Kaisey S. Mandel$^{1,3}$\orcidlink{0000-0001-9846-4417
}, Matthew Grayling$^{1}$\orcidlink{0000-0002-6741-983X},\newauthor Abhijit Saha$^{4}$\orcidlink{0000-0002-6839-4881}, Tim Axelrod$^{5}$\orcidlink{0000-0002-5722-7199}, Thomas Matheson$^{4}$\orcidlink{0000-0001-6685-0479}, Edward W. Olszewski$^{5}$\orcidlink{0000-0002-7157-500X},\newauthor Annalisa Calamida$^{6}$\orcidlink{0000-0002-0882-7702}, Aaron Do$^{1}$\orcidlink{0000-0003-3429-7845}, Ralph C. Bohlin$^{6}$\orcidlink{0000-0001-9806-0551}, Jay B. Holberg$^{7}$\orcidlink{0000-0003-3082-0774}, Ivan Hubeny$^{5}$\orcidlink{0000-0001-8816-236X}, \newauthor
Susana Deustua$^{6,8}$\orcidlink{0000-0003-2823-360X},
Armin Rest$^{6,9}$\orcidlink{0000-0002-4410-5387},
Christopher W. Stubbs$^{10,11}$\orcidlink{0000-0003-0347-1724}, Aidan Berres$^{2}$\orcidlink{0000-0002-5010-441X}, 
Mai Li$^{2}$\orcidlink{0009-0009-4858-4029},\newauthor 
John W. Mackenty$^{6}$\orcidlink{0000-0001-6529-8416} and Elena Sabbi$^{12,4,5,6}$\orcidlink{0000-0003-2954-7643}
\\
$^{1}$Institute of Astronomy and Kavli Institute for Cosmology, University of Cambridge, Madingley Road, Cambridge, CB3 0HA, UK\\
$^{2}$University of Illinois at Urbana-Champaign, 1002 W. Green Street, Urbana, IL 61801, USA \\
$^{3}$Statistical Laboratory, DPMMS, University of Cambridge, Wilberforce Road, Cambridge, CB3 0WB, UK\\
$^{4}$NSF’s National Optical Infrared Astronomy Research Laboratory, 950 North Cherry Avenue, Tucson, AZ 85719, USA\\
$^{5}$The University of Arizona, Steward Observatory, 933 North Cherry Avenue, Tucson, AZ 85721, USA \\
$^{6}$Space Telescope Science Institute, 3700 San Martin Drive, Baltimore, MD 21218, USA\\
$^{7}$The University of Arizona, Lunar and Planetary Laboratory, 1629 East University Boulevard, Tucson, AZ 85721, USA\\
$^{8}$Sensor Science Division, National Institute of Standards and Technology, Gaithersburg, MD 20899-8441, USA\\
$^{9}$Department of Physics and Astronomy, Johns Hopkins University, Baltimore, MD 21218, USA\\
$^{10}$Harvard University, Department of Physics, 17 Oxford Street, Cambridge, MA 02138, USA\\
$^{11}$Harvard-Smithsonian Center for Astrophysics, 60 Garden Street, Cambridge, MA 02138, USA \\
$^{12}$Gemini Observatory, 950 N. Cherry Avenue, Tucson, AZ 85719, USA\\
}

\date{Accepted XXX. Received YYY; in original form ZZZ}
\pubyear{2024}

\usepackage{array}

\usepackage{orcidlink}

\begin{document}
\renewcommand{\arraystretch}{1.25} 
\label{firstpage}
\pagerange{\pageref{firstpage}--\pageref{lastpage}}
\maketitle

\begin{abstract}
We use hierarchical Bayesian modelling to calibrate a network of 32 all-sky faint DA white dwarf (DA WD) spectrophotometric standards ($16.5 < V < 19.5$) alongside three CALSPEC standards, from 912 \r{A} to 32 $\mu$m. The framework is the first of its kind to jointly infer photometric zeropoints and WD parameters (surface gravity $\log g$, effective temperature $T_{\text{eff}}$, extinction  $A_V$, dust relation parameter  $R_V$) by simultaneously modelling both photometric and spectroscopic data. We model panchromatic Hubble Space Telescope Wide Field Camera 3 (HST/WFC3) UVIS and IR photometry, HST/STIS UV spectroscopy and ground-based optical spectroscopy to sub-percent precision. Photometric residuals for the sample are the lowest yet yielding $<0.004$ mag RMS on average from the UV to the NIR, achieved by jointly inferring time-dependent changes in system sensitivity and WFC3/IR count-rate nonlinearity. Our GPU-accelerated implementation enables efficient sampling via Hamiltonian Monte Carlo, critical for exploring the high-dimensional posterior space. The hierarchical nature of the model enables population analysis of intrinsic WD and dust parameters. Inferred spectral energy distributions from this model will be essential for calibrating the James Webb Space Telescope as well as next-generation surveys, including Vera Rubin Observatory's Legacy Survey of Space and Time and the Nancy Grace Roman Space Telescope.
\end{abstract}

\begin{keywords}
methods: statistical --  white dwarfs -- standards -- surveys -- stars: statistics -- stars: fundamental parameters \vspace{-0.3cm}
\end{keywords}



\section{Introduction}

As we approach a new era of precision cosmology, our systematic uncertainties will outweigh our statistical uncertainties. This effect will be further amplified with the next-generation of wide-field surveys, such as the Legacy Survey of Space and Time at the Vera Rubin Observatory \citep[LSST;][]{lsst1}, as Type Ia supernova (SN Ia) sample sizes increase to over a million.

A particularly delicate systematic uncertainty to consider is the spectrophotometric calibration of surveys \citep{stubbs2015}. Calibration can be absolute, aiming to define the flux of an object in relation to International System of Units \citep{bohlin2014b,gordon2022}. Alternatively, the calibration can be relative, where an arbitrary flux scale is chosen, but the objective is to characterise an object's change in flux as a function of wavelength \citep{n19,a23}. In this work, we introduce \textit{DAmodel}: a hierarchical Bayesian framework for accurately calibrating the relative flux of 32 faint DA white dwarfs, along with three primary standards. The model utilises photometric and spectroscopic data from diverse sources, achieving unprecedented levels of flux calibration, ready for the demands of next-generation surveys.

 There is a particularly strong degeneracy between the photometric redshift of a galaxy and the relative calibration of a survey's photometric zeropoints \citep{high2009,euclid2024}. If a survey's relative zeropoints are calibrated incorrectly, the difference in flux could be misinterpreted as additional colour that would then compromise the photometric redshift estimate. This effect can also influence the Hubble diagram in SN Ia cosmology. Although systematic effects only change colours of individual objects by the order of a percent \citep{betoule13}, the combined effect of incorrectly measuring an entire SN population's colour would have a significant influence on cosmological constraints \citep{hounsell2018}. \cite{scolnic2014} showed that colour calibration systematics can create an additional signal on the Hubble diagram at the 0.01-0.015 magnitude level. In the DESI analysis \citep{desi2024}, the difference between an evolving $w(z)$ and constant $w=-1$ dark energy equation of state universe was a signal of only 0.04 mag, emphasising the importance of having confidence in relative photometric calibration. A further process in supernova cosmology that requires accurate relative flux calibration is the comparison of low redshift and high redshift samples. These contrasting samples are necessary to measure the change in the relative curve of the Hubble diagram, however, if these samples use inconsistent definitions of brightness or colour then they will be at odds with our true underlying cosmology \citep{efstathiou2024,dhawan2024}. To overcome this issue, it is imperative to perform reliable cross-calibration between surveys that rely on accurate flux standards.

\begin{figure*}
\label{fig:globe}
\includegraphics[width=1.8\columnwidth]{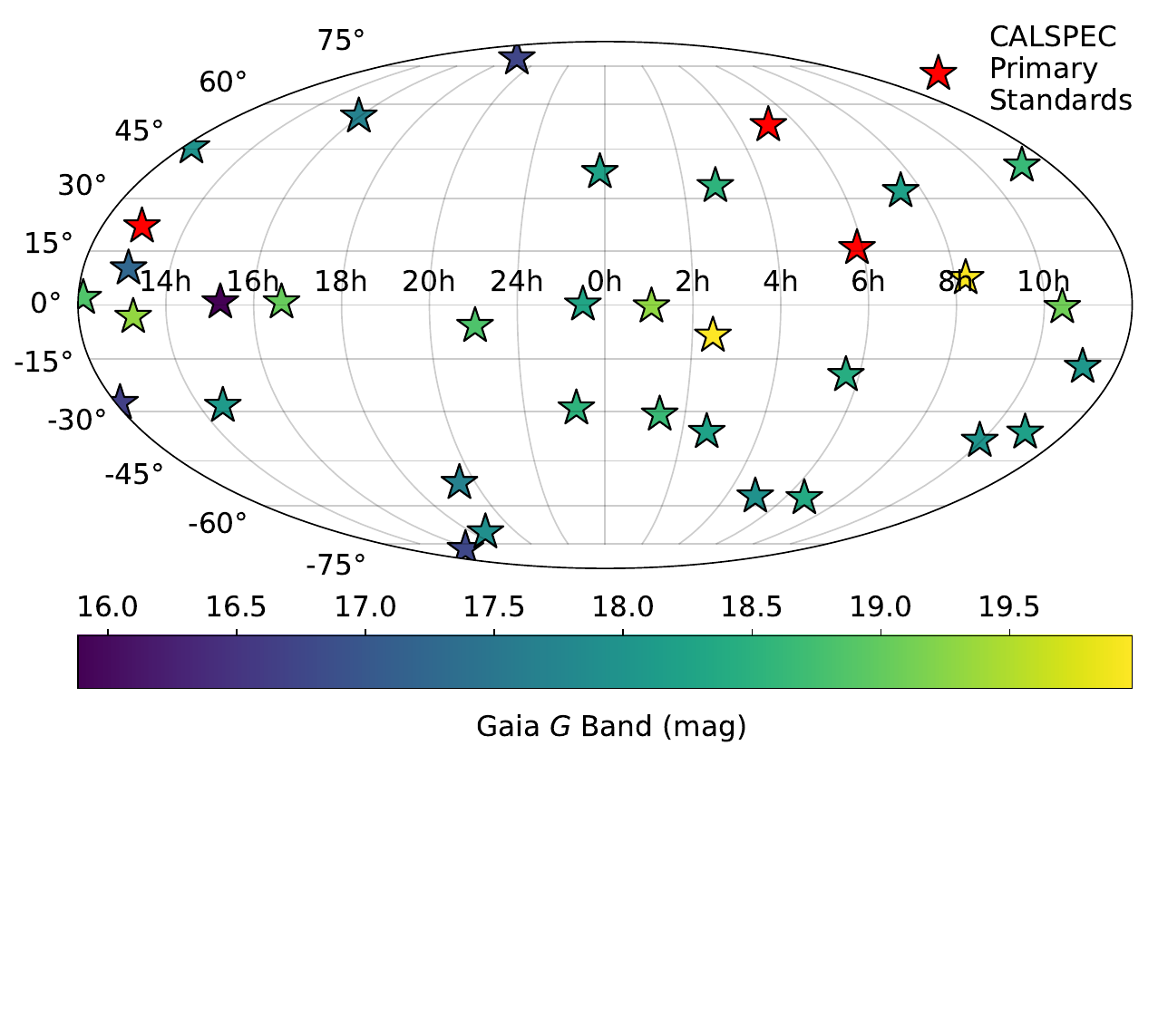}\vspace{-3.8cm}
 \caption{Coordinates and \textit{Gaia} $G$ band brightness of the DA white dwarfs in our network. The red stars illustrate the CALSPEC primary standards that have brightnesses 11.7< $G$ <13.3. }
    \label{fig:globe_figure}
\end{figure*}

Spectrophotometric calibration relies on the precise modelling of  the spectral energy distributions (SEDs) of astronomical objects that are well understood. Pure-hydrogen atmosphere DA white dwarfs (DA WDs) are ideal candidates to be modelled as standards, since they have simple continuum-dominated spectra and have well characterised Balmer absorption features that are correlated with surface gravity and effective temperature \citep{holberg1985,holberg2006,tlusty}. So far, the three most accurate DA WDs used for spectrophotometric calibration have been GD71, GD153 and G191B2B. These WDs have little reddening and have been staples of the CALSPEC system \citep{bohlin2014a,bohlin2020}, allowing for the calibration of astronomy's most successful space missions, such as the \textit{Hubble Space Telescope} (HST) and \textit{Gaia} \citep{gaia2016}, as well as various other ground-based instruments. Since different missions scan different regions of the sky, it is necessary for networks of standards to have coverage over large portions of the celestial sphere. 
\cite{rubin2022prop} was awarded 35 HST/STIS orbits to observe 11 DA WDs to be added to CALSPEC system. \cite{elms2024} recently expanded the CALSPEC standards by adding 17 new cool ($T_{\text{eff}}$  < 20~000 K) DA WDs. Low temperature DA WDs are  particularly useful for standardisation as they have reduced non-local thermal equilibrium effects (more predictable SEDs) and are brighter in the near-infrared (NIR).

An essential factor in spectrophotometric calibration that must be treated with care is the modelling of the effects of interstellar dust within the Milky Way. Dust along the line of sight between an object and the Earth has the combined effect of extinguishing and reddening light \citep{driane2003}. Extinction, often measured in the $V$ band as $A_V$, is dependent on the column density of dust that lies between object and observer \citep[e.g.][]{foster2013}, which is also correlated with distance. The quantity $R_V=A_V/E(B-V)$, where $E(B-V)$ is the colour excess between $B$ band and $V$ band, is used to determine the amount of reddening one would expect from different types of dust grains. There have been various functional forms of dust relations that use these parameters to define extinction as a continuous function of wavelength $A(\lambda; A_V,R_V) $ \citep[e.g.][]{cardelli1989,Fitzpatrick1999,fitzpatrick19,gordon2021,gordon2023,decleir2022}, that have been a key ingredient in spectrophotometric modelling DA WDs. Since DA WDs are so well understood, they are also useful tools to study dust. \cite{sahu2024} recently used DA WDs to validate three-dimensional dust maps in the UV.

The value of $A_V$ has a strong impact in the UV wavelengths, particularly around the 2175 \AA \space bump that leads to a sharp drop off in extinction \citep{fm1986}. For this reason, it is important to spectroscopically observe DA WDs in the UV to constrain $A_V$ and extrapolate SEDs with a higher degree of confidence to longer wavelengths in the thermal and far infrared. Dust extinction decreases significantly in the NIR, meaning that if we have data coverage at these wavelengths we can disentangle degenerate dust and intrinsic parameters. The difference in extinction at UV and NIR wavelengths allows for more reliable inference of the $R_V$ parameters. This panchromatic strategy to mitigate dust effects is also used elsewhere in astronomy, for example in modelling SNe Ia (Type Ia Supernovae) \citep{mandel2020,thorpmandel22,grayling2024,thorp2024,grayling_popivic2024} and inference of galaxy properties \cite{alsing2024,thorp2024_pop}.

As we look towards the next-generation of deep surveys such as LSST \citep{lsst1}, the \textit{Nancy Grace Roman Space Telescope }\citep[Roman;][]{roman2012} and the \textit{James Webb Space Telescope} \citep[JWST;][]{jwst2006}, we need dimmer standards to calibrate them. The three primary CALSPEC DA WDs, with magnitudes ranging from 11.7 < $V$ < 13.4, have been observed with JWST but are close to its observational limit and exceed the normal observing mode saturation limits of upcoming surveys, motivating the need for a new set of fainter standards. Our collaboration (initiated and led by co-author Saha) has established an all-sky network of 32 faint (16.5 < $V$ < 19.5) DA white dwarfs to be used as spectrophotometric standards \citep{n16,n19,c19,calamida2022,a23}.  This group of DA WDs, illustrated in Figure \ref{fig:globe}, is intended to deliver exceptionally precise flux calibration, ready for the upcoming era of deep-sky surveys. 

\cite{n19} developed a novel Bayesian methodology to model the faint WDs using HST photometry paired with ground-based optical spectroscopy \citep{c19} that would go on to achieve sub-percent accuracy. The approach consisted of a two-step process. First, HST/WFC3 photometry was calibrated for each WD across six bands, spanning from the UV to the NIR. This step included addressing various instrumental systematics, such as time-dependent changes in zeropoint sensitivity in each band \citep{calamida2022zp,marinelli2024}. The second step of the process took the calibrated photometry and ground based optical spectra to serially model SEDs of each WD by inferring intrinsic and dust parameters. 

The work of \cite{n19} was particularly successful in the optical and UV wavelengths, yielding an average RMS of 4 mmag across the five bands. The NIR F160W band (centred at 1537 nm) proved to be more difficult due to known count-rate nonlinearity (CRNL) systematics \citep{riess2019,bohlin2019}. \cite{a23} subsequently expanded and reanalysed the network by using the approach developed by \cite{n19} to establish spectrophotometric priors on the model parameters. They then transitioned to a hierarchical Bayesian model to address the CRNL more accurately than using photometry alone. This combination proved successful, accomplishing a RMS of 8 mmag in the NIR, demonstrating the benefit of hierarchically analysing the network. The hierarchical approach enabled the inference of a band-dependent zeropoint shift relative to the CALSPEC system, which can only be derived from a set of stars. This allowed relative independence from CALSPEC which was crucial for ensuring the network could be calibrated to any magnitude system. It became clear that the next objective for this network would be a fully hierarchical end-to-end Bayesian analysis that would combine all spectroscopic and photometric data available.

Hierarchical Bayesian models (HBMs) have become a powerful tool in astronomy, offering a flexible way to handle complex functional data with multiple layers of uncertainty \citep[e.g.][]{loredo2024}. Their strength lies in the ability to model dependencies across different levels, such as population parameters for groups of objects and individual-level parameters that vary within these groups. This hierarchical approach enables astronomers to account for intrinsic scatter, observational uncertainties, and selection biases, which are common in astronomical surveys. HBMs have been applied across many areas of astronomy for a variety of applications, including exoplanet atmospheric retrievals \citep{exohbm}, stellar evolution \citep{shijing2017,shijing2018}, galaxies \citep{loredo2019,leistedt2023}, gravitational wave analysis \citep{gwhbm}, SN Ia modelling \citep{mandel2009,mandel2011,mandel2014,mandel2017,mandel2020} and cosmology \citep{march2011,rubin2015,feeney2018,boyd2024}. A HBM was used in \cite{rubin2022} to perform a simultaneous spectrophotometric recalibration of 32 stars, including 14 existing CALSPEC standards, based on measurements obtained with the SuperNova Integral Field Spectrograph. The work was able to achieve 14 mmag accuracy by carefully modelling population-level instrumental systematics relating to atmospheric extinction, temperature dispersion and nightly calibration. In addition, HBMs were also used for the calibration of X-ray surveys in \cite{chen2019} and \cite{marshall2021}. 

In this work we establish \textit{DAmodel}: a scalable fully hierarchical Bayesian framework for spectrophotometric calibration of our all-sky DA WD network (including 32 faint and 3 CALSPEC standards) to the highest accuracy yet, allowing for precision astronomy in the next-generation of surveys. For the first time, we jointly analyse HST/WFC photometry in five UVIS broad bands and one NIR band, high resolution optical ground-based spectra \citep{c19,calamida2022} and new UV spectra from the Hubble Space Telescope Imaging Spectrograph (STIS). This work solves the calibration problem by implementing a novel multi-level hierarchy that can use the network to simultaneously infer instrumental systematics, dust relations and intrinsic WD parameters. 

In Section \ref{sec:data}, we outline the spectrophotometric DA WD dataset used in this work. We then outline the statistical model of the \textit{DAmodel} framework in Section \ref{sec:stats}. In Section \ref{sec:results} we present our calibrated SEDs, synthetic photometric residuals with HST and other surveys, then discuss the implications of our results. Future applications of the framework are discussed in Section \ref{sec:future} and we conclude with final remarks in Section \ref{sec:concl}.
\begin{figure*}
\label{fig:data_plot}
\includegraphics[width=2\columnwidth]{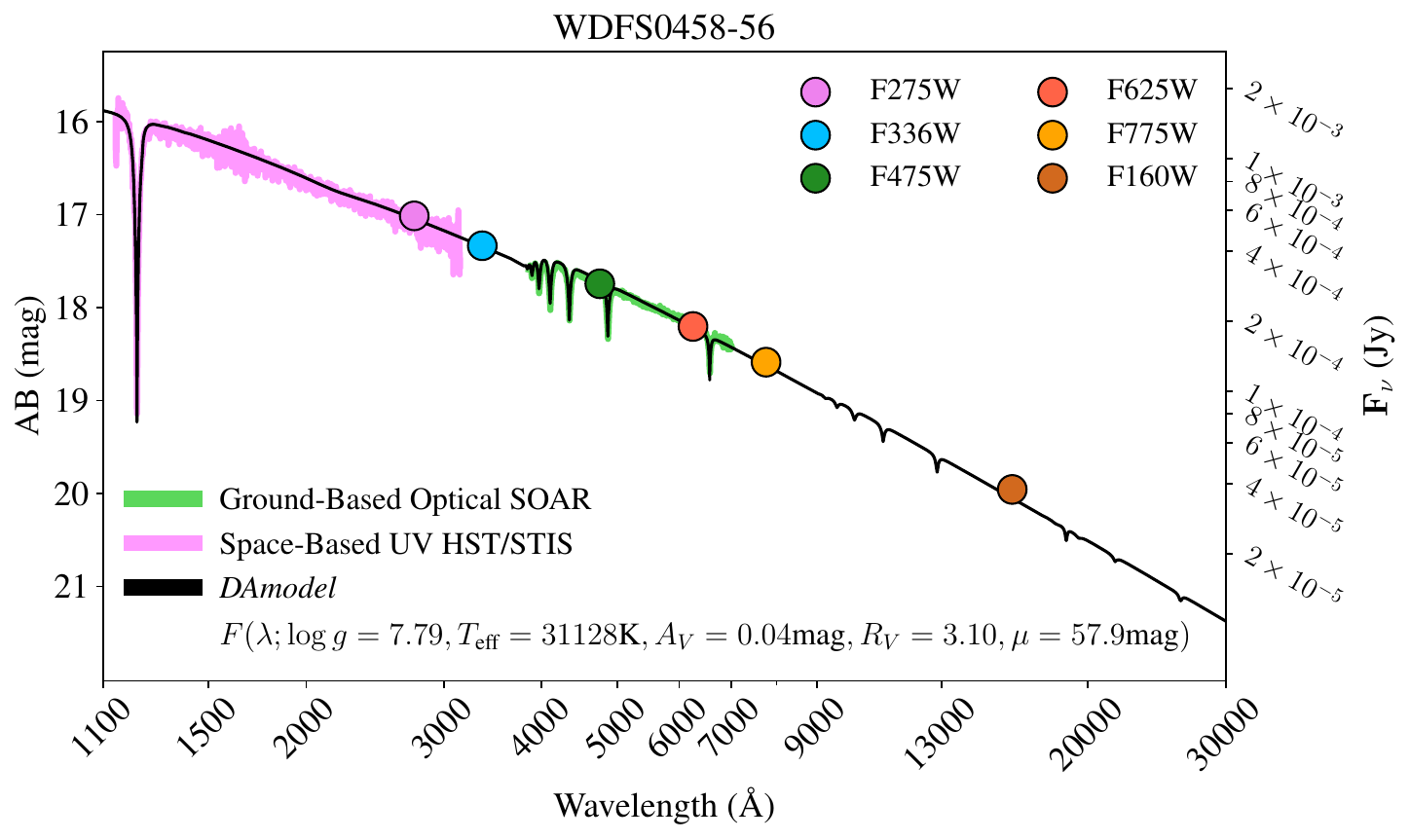}
\caption{Visualisation of available data for WDFS0458-56. The coloured circles represent HST/WFC3 UVIS and IR photometry, increasing in wavelength from left to right. Highlighted in green is the ground-based optical spectrum from the Southern Astrophysical Research (SOAR) telescope. Shaded in pink is the new HST/STIS UV spectrum that we have also obtained for 18 other faint DA WD sources. Under-plotted in black is our SED model evaluated at the parameters inferred for this particular source.  }
    \label{fig:datapplot}
\end{figure*}
\section{Data}
\label{sec:data}

All the data used in this work has been previously described and analysed individually in other publications, however, this is the first time in which all the data has been modelled simultaneously. The northern and equatorial standards were analysed in \cite{n19} using optical spectra and photometry detailed in \cite{c19}. The photometry and optical spectra of the full network was analysed in \cite{a23} and was introduced in \cite{calamida2022}. More recently, the HST UV spectra for a subset of the network was published and analysed in \cite{bohlin2025stis}. In this section we briefly describe the DA WD network and the observational data used in our work.

\subsection{Target Selection}
\label{sec:targets}

\cite{n19}  and \cite{c19} selected 19 northern and equatorial faint DA WDs using known sources, while \cite{calamida2022} established an additional 13 southern standards using their own methodology. Standards were carefully selected to avoid for any factors such as magnetic line splitting or trace atmospheric elements in their spectra, as these effects were not included in our modelling. Standards were also monitored to ensure there was no photometric variability. Hot DA white dwarfs are not expected to vary intrinsically, but variability could arise due to binary companions, “seeing” effects from nearby faint red stars or dust clouds surrounding the white dwarf. Absolute brightnesses of the stars were chosen to be no brighter than $V=9.0$. Only DA WDs with effective temperatures above 19~500~K were kept to be within the bounds where we have confidence in theoretical templates. The final all-sky set of 32 faint DA WDs that met the selection criteria was published in \cite{calamida2022}. These are combined with the three CALSPEC primary standards to make the full network of 35 stars that we analyse in our work. Effective temperatures of the network range from 20~000~K to 67~000~K and the natural logarithm of surface gravities are between 7.2 and 9.0. The \textit{Gaia} $G$ band apparent brightnesses of the faint standards range between 15.9 mag and 20.0 mag. Figure \ref{fig:globe} illustrates the locations and brightnesses of our faint all-sky network. Coordinates of the WDs can be found in Table \ref{tab:wd_params} and online at \url{https://zenodo.org/records/14939532/files/wdfs_coords.txt}.

\subsection{Hubble Space Telescope WFC3 Photometry}
\label{sec:phot_data}

The optical photometry used in this study is extensively documented in \cite{c19} and \cite{calamida2022}, while here we summarise  important details for our modelling. The photometry was acquired with the WFC3 UVIS and IR channels during Cycle 20 (average MJD=56452.94 days), Cycle 22 (average MJD=57076.11 days) and Cycle 25 (average MJD=57588.79 days) with programs  GO-12967, GO-13711 and GO-15113 respectively (PI Abhijit Saha). Cycle 20 observed six of the northern and equatorial faint standards in  five HST filters: F336W, F475W, F625W, F775W, and F160W. Cycle 22 observed all 19 of the northern and equatorial faint standards, as well as the three CALSPEC primary standards, in the same five filters and the additional near-UV F275W filter. Cycle 25 observed the 13 southern standards and the three primary CALSPEC standards using all six bands from the near-UV to the near-infrared. The sensitivity of WFC3 has changed as a function of time \citep{calamida2022zp}, so we can use the differences of the faint standards observed in both Cycle 20 and 22, as well as change in the CALSPEC primary standards in Cycle 22 and 25, to help constrain the varying zeropoints.

The bands in the near-UV and optical were on different UVIS sub-arrays in each cycle. Observations in Cycle 20 used the UVIS1 sub-arrays and remained on the UVIS1 photometric system, while Cycle 22 observations used UVIS2 sub-arrays and were on UVIS2 photometric system. Observations in Cycle 25 were taken with the UVIS2 sub-arrays but converted back to the UVIS1 photometric system in the \textit{calwf3} calibration pipeline. The differences in photometric system are important to consider when modelling the changes in zeropoints  between cycles. The final photometry was placed on the AB magnitude system of CALSPEC. Although the use of AB magnitudes is not critical for subsequent analysis, it provides a convenient and self-consistent reference system.

\subsection{Ground-Based Optical Spectra}
\label{sec:spec_data}

Ground-based optical spectra were obtained with the Gemini North and South telescopes, the Multiple Mirror Telescope (MMT) and Southern Astrophysical Research (SOAR) Telescope, as described in \cite{c19} and \cite{calamida2022}. The observations of the standards with Gemini spectrographs ($1.0''$ to $1.5''$ slit) cover a continuous range from 3500 \AA\space to 6360 \AA. The MMT observations of the standards ($1.0''$ to $1.25''$ slit) cover a wavelength range from 3400~\AA \space to 8400~\AA. The DA WDs were observed with SOAR ($1.07''$ slit) in a wavelength range between 3850~\AA\space and 7100~\AA. The spectrographs did not operate on a linear pixel-to-angstrom scale, so the data was resampled onto a uniform linear grid after extraction, with a sampling of 1 \AA/pixel for Gemini, 2 \AA/pixel for MMT and 2 \AA/pixel for SOAR. For objects with multiple spectroscopic observations in the optical, we default to using the same spectrum used in the \cite{n19} and \cite{a23} analyses.

\subsection{Space-Based STIS UV Spectra}
\label{sec:stis_data}
To better characterise and ultimately correct for dust effects, the spectroscopic observations were extended into the UV regime. The HST Space Telescope Imaging Spectrograph (STIS) was used to observe 19 of the faint DA WDs in program 16764 (PI Gautham Narayan). These stars had apparent magnitudes in the F275W filter between 15 mag and 18~mag. Spectra were composed of observations using both FUV and NUV detectors with the G140L and G230L gratings, covering a combined wavelength range of 1150~\AA \space to 3180~\AA. The resolution was resampled to 1 \AA/pixel for our analysis. The specific WDs that had this additional coverage are indicated in Table \ref{tab:wd_params} and Table \ref{tab:synth_phot}. Further information on the STIS UV spectra can be found in the preliminary analysis by \cite{bohlin2025stis}.

\begin{figure*}
	\includegraphics[width=1.8\columnwidth]{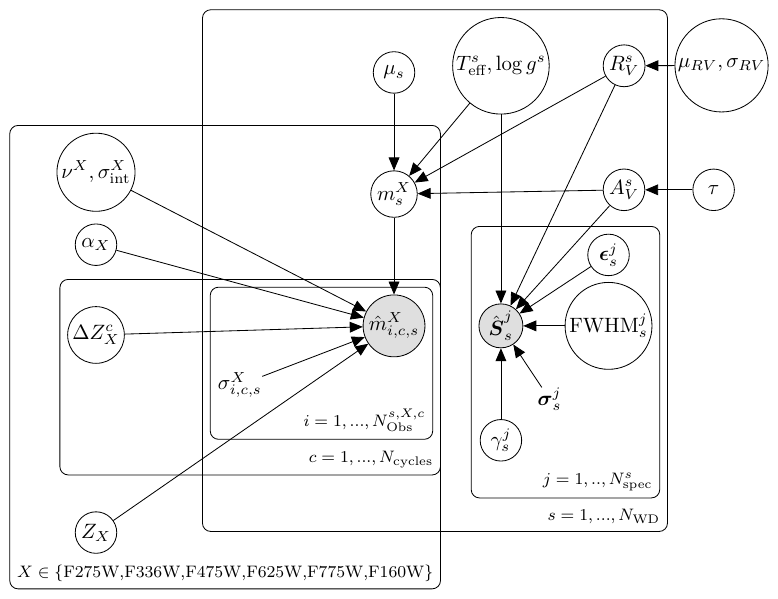}
    \caption{Probabilistic graphical model (directed acyclic graph) illustrating the relationships between parameters in our hierarchical Bayesian model. In the diagram we see multiple levels of of hierarchy where the top level population dust parameters ($\mu_{RV}$,$\sigma_{RV}$,$\tau$) influence each WD's extinction. The object-level parameters ($\mu_s$,$T_{\text{eff}}^s$, $\log g^s$,$A_V^s$,$R_V^s$) determine the SED of each individual WD. The nuisance parameters ($\bm{\epsilon}^j_s$,$\text{FWHM}^j_s$,$\gamma^j_s$) allow us to model spectra $\bm{\hat{S}}^j_s$ from the each object's SED. There are systematic parameters ($Z_X$,$\alpha_X$,$\sigma^X_{\text{int}}$,$\nu^X$) in each band $X$, as well as HST cycle-dependent offsets $\Delta Z^c_X$, allowing us to model photometric observations $\hat{m}_{i,c,s}$ of the network. Each parameter is defined clearly in Section \ref{sec:stats}.}
    \label{fig:graphical_model}
\end{figure*}
\section{The Statistical Model}
\label{sec:stats}
In this section we define the \textit{DAmodel} statistical model, building on techniques previously developed by \cite{n19} and \cite{a23}. In Section \ref{sec:sed} we describe the forward model to construct our dusty SEDs. In Section \ref{sec:phot} we describe how we use the dusty SEDs in our likelihoods to model the  HST/WFC photometry and instrumental systematics. In Section \ref{sec:spec} we describe how we adapt the dusty SEDs to model observed optical and STIS UV spectra. In Section \ref{sec:hbm} we describe how we combine the different elements of the statistical model to make a fully hierarchical Bayesian model, which is also illustrated in Figure \ref{fig:graphical_model}. Finally, in Section \ref{sec:priors} we define the choice of priors on each inferred parameter.

\subsection{Synthetic Spectral Energy Distribution}
\label{sec:sed}
The model requires an underlying grid to describe the relationship between the spectral energy distribution (SED) of the DA white dwarf with intrinsic parameters describing effective temperature $T_{\text{eff}}$ and the natural logarithm of the surface gravity $\log g$. The distinct depths of the Balmer series in the optical wavelengths are strongly correlated with $T_{\text{eff}}$, whereas their equivalent widths are dependent on  $\log g$ \citep{tremblay09}. 

\cite{n19} and \cite{a23} previously used a non-local thermodynamic equilibrium (NLTE) model grid computed by \texttt{Tlusty} v202 code \citep{tlusty}. In this work we use the recently updated NLTE grid\footnote{The updated theoretical model spectra for hot DA white dwarfs used in this work can be found at: \url{www.as.arizona.edu/~hubeny/pub/DAgrid24.tar.gz}} computed with \texttt{Tlusty} v208 \citep{hubeny2021} that provides us with improved modelling of Lyman-$\alpha$ at UV wavelengths, necessary for analysis of the STIS spectra. There are still future updates to come with the theoretical DA WD grids, particularly with development of hydrogen Lyman lines Stark broadening, as well as improvements in the UV for cooler DA WDs at $T_{\text{eff}}$ < 20~000 K. Our method is general and can seamlessly integrate future grids as they become available.

The new \texttt{Hubeny} grid is defined from 900 \AA \space to 32 $\mu$m with resolution of $R=$100~000 for DA WDs in the range of 15~000~K < $T_{\text{eff}}$ < 70~000 K  and 7 < $\log \text{g}$ < 9.5. The resolution of the grid is 0.25 for $\log \text{g}$ and 250 K, 500 K and 1000 K for effective temperatures up to 20~000 K, 40~000 K and 70~000~K respectively. Our model's unreddened SED $\mathcal{F}(\lambda;\log g , T_{\text{eff}})$ is determined by the cubic interpolation of the theoretical grid. The SED is then reddened according to the \cite{gordon2023} dust relation \citep{gordon2009,gordon2021,fitzpatrick19,decleir2022,gordon2024a,gordon2024b} where the extinction $A(\lambda;A_V,R_{V})$ is applied such that
\begin{equation}
\label{eq:redsed}
    \Tilde{F}(\lambda;\log g , T_{\text{eff}},A_V,R_V) = \mathcal{F}(\lambda;\log g , T_{\text{eff}}) \cdot 10^{-0.4 \cdot{A(\lambda;A_V,R_V})}
\end{equation}
where $A_V$ is the extinction in $V$ band and $R_{V}=A_V/E(B-V)$, where $E(B-V)$ is the colour excess between $B$ band and $V$ band.  The extinction relation is defined between 912 \AA\space and 32 $\mu$m which is what dictates the limits of our published SEDs. In Section \ref{sec:f99g23} we discuss how using \cite{gordon2023} dust relation over \cite{Fitzpatrick1999}, that was used in previous analyses, affects our results.

We apply an achromatic offset to the reddened SED using 
\begin{equation}
    F(\lambda;\log g , T_{\text{eff}},A_V,R_V,\mu) = \Tilde{F}(\lambda;\log g , T_{\text{eff}},A_V,R_V)  \cdot 10^{-0.4 \mu}
\end{equation}
The pseudo-distance modulus 
$\mu$ is a scale factor expressed in magnitudes with an arbitrarily defined zeropoint, relating an object's surface brightness to its apparent brightness. The value of 
$\mu$ depends on both the physical size of the object and its distance from the observer. In our analysis methodology, neither the physical size nor the distance is determined individually, as we do not impose a mass-radius-luminosity relation to derive the absolute brightness.

\subsection{Photometric Modelling}
\label{sec:phot}
To yield synthetic photometric magnitudes we integrate a WD's SED spectrum $F(\lambda;\bm{\Theta_{\text{WD}}},\mu)=F(\lambda;\log g , T_{\text{eff}},A_V,R_V,\mu)$ with the given HST/WFC3 passband $X\in \{$F275W, F336W, F475W, F625W, F775W, F160W$\}$ throughput response\footnote{UVIS2 and IR filters from \cite{calamida2022zp} available at: \url{https://www.stsci.edu/hst/instrumentation/wfc3/performance/throughputs}} $T_X(\lambda)$ such that 
\begin{equation}
\label{eq:flux_int}
    m_s^X = -2.5 \log_{10} \Bigg (\frac{\int^{\infty}_0 \lambda \cdot F(\lambda;\bm{\Theta^s_{\text{WD}}},\mu_s) \cdot T_X(\lambda) \cdot d \lambda}{\int^{\infty}_0 \lambda\cdot F_{\text{AB}}(\lambda)\cdot T_X(\lambda)\cdot  d \lambda}\Bigg)
\end{equation}
where $F_{\text{AB}}(\lambda)$ is the spectral flux density of the fiducial AB source expressed in $\text{ergs cm}^{-2}$ $\text{s}^{-1}$ $\text{\AA}^{-1}$. Intrinsic WD and dust parameters collectively represented by $\bm{\Theta^s_{\text{WD}}}=(\log g^s ,T^s_{\text{eff}},A^s_V,R^s_V)^T$.

We then construct a likelihood to model individual photometric observations $\hat{m}^X_{i,c,s}$ of WD $s$ in band $X$ from Hubble Cycle $c$ as
\begin{multline}
P\big(\hat{m}^X_{i,c,s}|\ \bm{\Theta^s_{\text{WD}}}, \mu_s,\bm{\theta_{\text{phot}}^X},\Delta Z_X^c\big)= \\
    \mathcal{T}\big(\hat{m}^X_{i,c,s}|\ m^X_s-Z_X-\Delta Z_X^c-\alpha_{X}(m^X_s-15),\sqrt{\sigma^{X\hspace{0.3cm}2}_{i,c,s}+\sigma_{\text{int}}^{X\hspace{0.1cm}2}},\nu^X \big)
\end{multline}
where $\mathcal{T}(x|\ \bar{\mu},\sigma,\nu)$ is a student-T distribution with mean $\bar{\mu}$, scale $\sigma$ and $\nu$ degrees of freedom. We choose a student-T distribution to model the photometry due to its broader tails, allowing flexibility and robustness in modelling outliers. The measurement error for each individual photometric observation $i$ is represented by $\sigma^X_{i,c,s}$. Each HST band $X$ has its own set of parameters $\bm{\theta_{\text{phot}}^X}=(Z_X,\sigma_{\text{int}}^{X},\nu^X,\alpha_X)^T$ containing a zeropoint $Z_X$, excess dispersion $\sigma_{\text{int}}^{X}$, degree of freedom  $\nu^X$ and count-rate nonlinearity (CRNL) coefficient $\alpha_X$ \citep{riess2019,bohlin2019,a23}. The excess dispersion parameter $\sigma_{\text{int}}^{X}$ accounts for any additional band-dependent systematic scatter that is not accounted for in the measurement error of each observed object. If there are no additional systematics and the measurement errors describe the data perfectly, the model will infer $\sigma_{\text{int}}^{X}$ to be zero. We only consider a CRNL coefficient in the F160W band and set the coefficients for the remaining bands to zero. There is also an additional zeropoint offset $\Delta Z_X^c$ accounting for changes in zeropoints between Hubble cycles \citep{n19,calamida2022zp,marinelli2024} and differences in photometric settings. For at least one Hubble Cycle we must set $\Delta Z_X^c=0$ to avoid degeneracy with $Z_X$. For our analysis we set $\Delta Z_X^{25}=0$, then infer $\Delta Z_X^{20}$, and $\Delta Z_X^{22}$.

In order to infer zeropoints we need to tie our network to a photometric system. In principle this means setting at least one magnitude in a single band to a constant value. For self consistency, we arbitrarily tie the three bright WDs to the \cite{bohlin2020} CALSPEC system by setting the F475W magnitudes of primary standards, G191B2B, GD153 and GD71 to constant. The primary standards have known calibrated magnitudes $\bar{m}^X_p$ that are calculated by integrating published SEDs\footnote{g191b2b\_mod\_012.fits,  gd153\_mod\_012.fits and  gd71\_mod\_012.fits from \cite{bohlin2020} available at: \url{https://archive.stsci.edu/hlsps/reference-atlases/cdbs/current_calspec/}}\citep{bohlin2020} with the HST/WFC3 filter responses. The likelihood is therefore only dependent on the photometry parameters $\bm{\theta_{\text{phot}}^X}$ and $\Delta Z_X^c$ such that
\begin{multline}
P\big(\hat{m}^X_{i,c,p}|\ \bm{\theta_{\text{phot}}^X},\Delta Z_X^c\big)= \\\mathcal{T}\big(\hat{m}^X_{i,c,p}|\ \bar{m}^X_p-Z_X-\Delta Z_X^c-\alpha_{X}(\bar{m}^X_p-15),\sqrt{\sigma^{X\hspace{0.3cm}2}_{i,c,p}+\sigma_{\text{int}}^{X\hspace{0.1cm}2}},\nu^X \big)
\end{multline}

\begin{figure*}
	\includegraphics[width=2.\columnwidth]{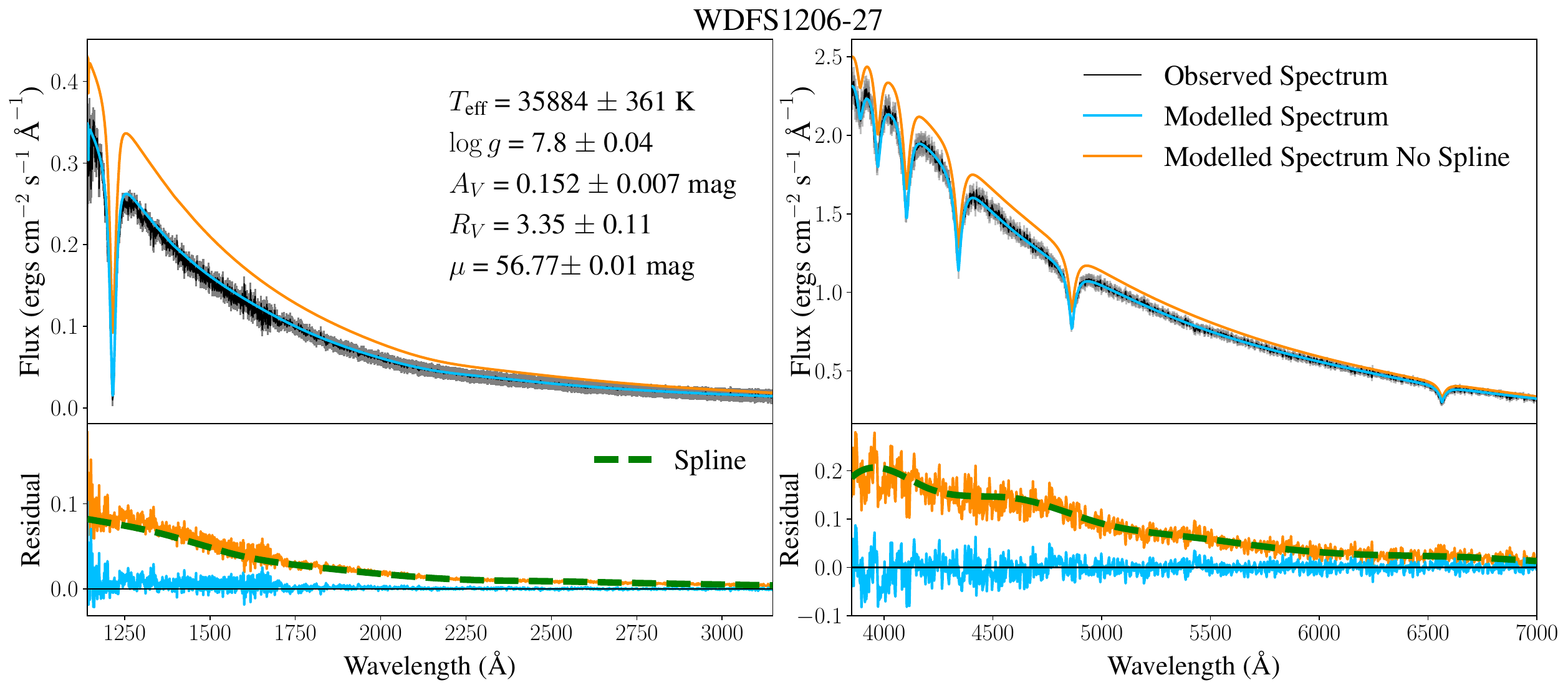}
 \vspace{-0.3cm}
    \caption{Example of modelled spectrum for WDFS1206-27. The black lines represent the observed spectra for the source from HST/STIS (left) and SOAR (right). The orange lines represent the inferred SED at the observed wavelengths. The green lines are the inferred cubic splines accounting for residual differences between the model and the data. These splines account for instrumental effects and inadequacies of the model, such as the dust relation or theoretical DA WD template. The blue line represents the modelled SED with the addition of the inferred cubic splines.   }
    \label{fig:spec_example}
\end{figure*}

\subsection{Spectroscopic Modelling}
\label{sec:spec}
For the spectroscopic modelling, we renormalise the reddened SED $\tilde{F}(\lambda;\bm{\Theta_{\text{WD}}}\big)=\tilde{F}(\lambda;\log g , T_{\text{eff}},A_V,R_V)$ defined in Equation \eqref{eq:redsed} and convolve it with a Gaussian kernel such that
\begin{multline}
f\big(\lambda; \bm{\Theta^s_{\text{WD}}},\gamma^j_{s},\text{FWHM}_s^j\big)
=\\ \frac{\tilde{F}(\lambda; \bm{\Theta^s_{\text{WD}}})}{4\pi \gamma_{j,s}^2} \hspace{0.1cm} \ast  \hspace{0.1cm} \mathcal{N}\big(\lambda|\ 0,\sigma_R(\text{FWHM}_s^j)^2\big)
\end{multline}
 The normalisation factor $\gamma^j_{s}$ acts as a nuisance factor to scale the synthetic spectrum to match observed spectrum $j$. We define $\mathcal{N}(x|\ 0,1^2)$ as a normal distribution with centring 0 and scale 1. The Gaussian kernel smooths the synthetic spectrum to be consistent with the observed spectrum's resolution, where absorption features are less sharp. The standard deviation of the kernel is defined as 
\begin{equation}
    \sigma_R(\text{FWHM}) = \frac{\text{FWHM}}{R\cdot \sqrt{8 \ln 2}}
\end{equation}
where $R$ is the resolution of the spectrum in \AA \space per spectral unit and $\text{FWHM}$ determines the extent of the smoothing. We adopt this simple and consistent smoothing approach because we are combining spectra from different ground and space-based instruments, each operating at different wavelengths with varying line-spread functions.

Further flexibility in the model is needed to correct for wavelength-dependent systematics caused by instrumental or calibration effects. This flexibility can also be thought of as inconsistencies between the model and the data, where either the dust relation or theoretical DA WD template disagrees with what is observed \citep{czekala2015}. In \cite{n19} the wavelength-dependent systematics were modelled using a Gaussian Process (GP; \citealt{rasmussen2005}) as an additive function. This approach allowed for maximum flexibility to learn the systematic but was also computationally expensive, involving large multivariate Gaussian covariances with the same dimensionality as the number of spectrum pixels. There was evidence in \cite{n19} (see Figure 9 of their work) that suggested the inferred function was simple and smooth, allowing us to choose a computationally cheaper method in our hierarchical work.

In our work we choose to model wavelength-dependent spectroscopic systematics using cubic splines. We take inspiration from \cite{mandel2020} by separating the cubic spline into a matrix $\bm{W}_s^j(\lambda;\bm{l}_s^j)$ of shape $N_{\text{pix}} \times M$, that is predetermined by the locations of spline knots $\bm{l}_s^j=\{l^{j,s}_1,...,l^{j,s}_{M}\}$, and a vector of spline coefficients $\bm{\epsilon}_s^j$ of shape $M \times 1$, such that
\begin{equation}
    \epsilon(\lambda;\bm{\epsilon}_s^j) = \bm{W}_s^j(\lambda;\bm{l}_s^j)\bm{\epsilon}_s^j
\end{equation}
We choose to place the $M=10$ spline knots equally between the minimum and maximum wavelength coverage of each observed spectrum $j$. The spline knot locations avoid absorption features and are far enough apart to not artificially add to their depths. We add the spline to the smoothed synthetic SED to make the modelled synthetic spectrum $S(\lambda;\bm{\Theta^s_{\text{WD}}},\bm{\theta}^{j,s}_{\text{spec}})$  using
\begin{equation}
    S(\lambda;\bm{\Theta^s_{\text{WD}}},\bm{\theta}^{j,s}_{\text{spec}}) =  f\big(\lambda; \bm{\Theta^s_{\text{WD}}},\gamma^j_{s},\text{FWHM}_s^j\big) +  \epsilon\big(\lambda;\bm{\epsilon}_s^j\big)
\end{equation}
where we gather all synthetic spectrum nuisance parameters into $\bm{\theta}^{j,s}_{\text{spec}}=(\gamma^j_{s},\text{FWHM}_s^j,\bm{\epsilon}_s^j)^T$. Figure \ref{fig:spec_example} demonstrates how including the smooth spline corrects systematics in modelled spectra and reduces residuals. 

We then construct the likelihood of the WD parameters given the observed spectrum $\bm{\hat{S}^j_s}$, representing the $j$th observation belonging to WD $s$, as
\begin{equation}
P(\bm{\hat{S}}^j_s|\ \bm{\Theta^s_{\text{WD}}},\bm{\theta}^{j,s}_{\text{spec}})=\prod_{k=1}^{K_s^j} \mathcal{N}\bigg(\hat{S}^{k,j}_s\bigg|\ S\big(\lambda^{s,j}_k;\bm{\Theta^s_{\text{WD}}},\bm{\theta}^{j,s}_{\text{spec}}\big),\sigma^{j,s\hspace{0.05cm}2}_{k}\bigg)
\end{equation}
We denote the wavelength pixel with $k$, where $\lambda^{s,j}_0$ is the first observed wavelength of spectrum $j$ measuring WD $s$ and $\lambda^{s,j}_K$ is the last. The observed flux value and corresponding error of the $k$th pixel in the observed spectrum are represented by $\hat{S}^{k,j}_s$ and $\sigma^{j,s}_{k}$. We make the assumption that each spectroscopic pixel $\hat{S}^{k,j}_s$ is conditionally independent given the WD parameters $\bm{\Theta^s_{\text{WD}}}$ and spectrum nuisance parameters $\bm{\theta}^{j,s}_{\text{spec}}$. Importantly the formulation of the spectroscopic modelling is robust to model multiple spectra $j$ per object with different wavelength coverage. In our work this feature is useful for inference using both optical and UV data, but in future work it is trivial to also include near-infrared spectra. For WDs that have multiple optical spectra we choose the spectrum used in \cite{n19}.

\subsection{Hierarchical Model}
\label{sec:hbm}
In constructing the hierarchical model, we make a series of assumptions that enable us to treat the random variables as independent. To begin with, we assume that intrinsic WD parameters are independent and that dust parameters are conditionally independent given their respective population hyperparameters. We assume that modelled photometry and modelled spectra for a given WD are conditionally independent given their intrinsic parameters and dust parameters. Each modelled spectrum is assumed to be conditionally independent given its WD parameters and has its own set of independent nuisance parameters. Modelled photometry for a given WD is conditionally independent given its photometric and cycle-dependent systematic parameters. We assume that all bands and cycles have their own set of systematics that are also independent. Finally, all observed photometric and spectroscopic measurement errors are assumed to be independent from each other and model parameters. The relationships are illustrated in the probabilistic graphical model in Figure \ref{fig:graphical_model}.

We combine the photometric and spectroscopic likelihoods into a combined hierarchical data likelihood for the full network at the same time, such that
\begin{multline}
P(\bm{\mathcal{D}}|\ \bm{\Theta}) = \\ \hspace{0.4cm}\prod_s^{N_{\text{WD}}}\Bigg[ 
\prod_X^{N_{\text{bands}}}\prod_c^{N_{\text{cycles}}}\prod_i^{N^{s,X,c}_{\text{Obs}}}P\big(\hat{m}^X_{i,c,s}\big|\ \bm{\Theta^s_{\text{WD}}}, \mu_s,\bm{\theta_{\text{phot}}^X},\Delta Z_X^c\big) \\ 
\times \prod_j^{N^{s}_{\text{spec}}}
P\big(\bm{\hat{S}}^j_s\big|\ \bm{\Theta^s_{\text{WD}}},\bm{\theta}^{j,s}_{\text{spec}}\big) \Bigg]
\end{multline}
where $\bm{\Theta}$ contains all white dwarf, spectroscopic and photometric parameters. We denote all spectroscopic and photometric data by $\bm{\mathcal{D}}$. 

We also define an additional likelihood term for the F475W band data $\bm{\mathcal{D}_p}$ of the $N_{\text{prim}}=3$ CALSPEC primary standards from \cite{bohlin2020} as 
\begin{multline}
P(\bm{\mathcal{D}_p}|\ \bm{\Theta}) = \\\prod_p^{N_{\text{prim}}}
\prod_c^{N_{\text{cycles}}}\prod_i^{N^{p,\text{F475W},c}_{\text{Obs}}} P\big(\hat{m}^\text{F475W}_{i,c,p}\big|\ \bm{\theta_{\text{phot}}^\text{F475W}},\Delta Z_\text{F475W}^c\big)
\end{multline}

The likelihood terms are combined with the priors on parameters using Bayes' rule to derive the posterior
\begin{equation}
P(\bm{\Theta}|\ \bm{\mathcal{D}}) \propto P(\bm{\mathcal{D}}|\ \bm{\Theta})P(\bm{\mathcal{D}_p}|\ \bm{\Theta}) P(\bm{\Theta})
\end{equation}
For sampling the hierarchical model we use Hamiltonian Monte Carlo \citep{duane1987}, a technique that uses gradient information (something that neural network architectures also use) for faster convergence on the posterior. We use the No-U-Turn Sampler (NUTS; \citealt{hoffman2011}) in \textit{numpyro} \citep{pyro, numpyro}, a probabilistic programming package for automatic differentiation and just-in-time (JIT) compilation built with \textit{JAX} \citep{jax2018github}. Using this software combination on a GPU node, the implementation takes approximately 30 minutes in total to produce four chains of 500 burn-in samples and 500 posterior samples, amounting to an average effective sample size of 1277. This is a significant speed-up compared to the previous implementation of the non-hierarchical model from \cite{n19} that would take roughly one CPU day per object to converge. When assessing the quality of our chains we diagnose convergence by confirming the Gelman-Rubin $\hat{R}$ statistic is less than 1.02 for each of the 3653 inferred parameters \citep{gelman1992,vehtari2021}. We also find that none of the chains result in divergent transitions \citep{betancourt2014,Betancourt2017}. The implementation of \textit{DAmodel} can be found at: \url{https://github.com/benboyd97/DAmodel}.

\subsection{Priors}
\label{sec:priors}
In our modelling we use priors $P(\bm{\Theta})$ on our parameters. The priors on intrinsic WD and dust parameters are defined as
\begin{alignat}{2}
P(\log g^s)&= \mathcal{U}(\log g^s |\ 7\text{}, 9.5\text{  }) \\
P(T_{\text{eff}}^s)&=\mathcal{U}(T_{\text{eff}}^s|\ 15000 \text{ K },70000 \text{ K })\\
P(R_{V}^S|\ \mu_{RV},\sigma_{RV})&= \mathcal{TN}(R_{V}^s|\ \mu_{RV},\sigma_{RV}^2,1.2,\infty)
\end{alignat}
\begin{equation}
P(A_V^s|\ \tau) =
  \begin{cases}
      \frac{1}{\tau} \exp\Bigg(-\frac{A_V^s}{\tau}\Bigg) & \text{if $A^s_V\geq $ 0}\\
      0 & \text{if $A^s_V < $ 0}\end{cases}
\end{equation}
We define $\mathcal{TN}(x|\ 0,1^2,b_l,b_u)$ as a truncated normal distribution with centring 0, scale 1, lower bound $b_l$ and upper bound $b_u$. The true mean and variance of a truncated normal distribution is dependent on the extent to which it is truncated. Our main results assume $\tau=0.1$ mag, $\mu_{RV}=3.1$ and $\sigma_{RV}=0.18$ motivated by \cite{schlafly2016}.

The priors on our population dust parameters when we infer them are
\begin{alignat}{2}
P(\mu_{RV})&=\mathcal{U}(\mu_{RV}|\ 1.2,\infty)
\\
P(\sigma_{RV})&=\mathcal{HC}(\sigma_{RV}|\ 0,1)\\
P(\tau) &= \mathcal{HC}(\tau|\ 0,1)
\end{alignat}
where $\mathcal{HC}(0,1)$ is a Half-Cauchy distribution with a scale of 1. The lower bound on $R_{V}$ of 1.2 is motivated by the Rayleigh scattering limit \citep{driane2003}.

The priors on $\mu_s$ and photometric parameters are defined as
\begin{alignat}{2}
P(\mu_s) &= \mathcal{U}(\mu_s|\  50,65)\\
P(\sigma_{\text{int}}^X)&=\mathcal{HC}(\sigma_{\text{int}}^X|\ 0,1)\\
    P(Z_X) &= \mathcal{N}(Z_X|\ \mu_Z^{X},1^2)\\
    P(\Delta Z_X^c) &= \mathcal{N}(\Delta Z_X^c |\  0,1^2) \\
    P(\nu^X) &= \mathcal{HC}(\nu^X|\ 0,5)\\
    P(\alpha_X) &= \mathcal{N}(\alpha_X|\ 0,0.01^2)
\end{alignat}
The $\mu_Z^X$ parameter is determined by taking the average difference between the measured (Cycle 25 only) and known calibrated magnitudes for the three CALSPEC standards for each band.

The spectroscopic nuisance parameter priors are defined as 
\begin{alignat}{2}
P(\gamma_s^j) &= \mathcal{U}(\gamma_s^j|\  0,20000)\\
P(\text{FWHM}^j_s) &= \mathcal{TN}(\text{FWHM}^j_s|\ 0,8^2,0,50)\\
P(\epsilon^j_s) &= \mathcal{N}(\epsilon^j_s|\ 0,1^2)
\end{alignat}

\begin{table*}
	\centering
	\caption{White dwarf and dust parameters inferred using \textit{DAmodel}. Parameters for stars with an $*$ were inferred from STIS UV spectra as well as ground based optical spectra. Results shown are the averages and standard deviations from our chains. Note that the achromatic offset $\mu$ cannot be treated exactly as distance modulus, as as the absolute magnitude of each WD is not known without imposing a mass-radius-luminosity relation. This table is available in the online supplementary material.}
	\label{tab:wd_params}
	\begin{tabular}{lccccccc} 
		\hline
		Object &R.A. &Dec.&$T_{\text{eff}}$ & $\log g$ & $A_V$&$R_V$&$\mu$\\
  &hh:mm:ss.s&dd:mm:ss.s&K & &mag &&mag\\
		\hline
G191B2B&05:05:30.61&+52:49:51.96&$61941\pm519$& $7.67\pm0.03$& $0.0\pm0.0$& $3.1\pm0.18$& $52.59\pm0.01$ \\ 
GD153&12:57:02.34&+22:01:52.68&$41437\pm276$& $7.76\pm0.03$& $0.02\pm0.0$& $3.18\pm0.15$& $53.74\pm0.01$ \\ 
GD71&05:52:27.61&+15:53:13.75&$33719\pm118$& $7.88\pm0.01$& $0.02\pm0.0$& $3.24\pm0.16$& $53.13\pm0.01$ \\ 
\hline
WDFS0103-00&01:03:22.201&-00:20:47.800&$64833\pm3796$& $7.55\pm0.06$& $0.13\pm0.01$& $2.94\pm0.16$& $60.06\pm0.05$ \\ 
WDFS0122-30$^*$&01:22:00.725&-30:52:03.95&$36474\pm628$& $7.66\pm0.05$& $0.08\pm0.01$& $3.08\pm0.16$& $58.86\pm0.02$ \\ 
WDFS0228-08&02:28:17.183&-08:27:16.301&$22597\pm483$& $7.96\pm0.03$& $0.14\pm0.02$& $3.11\pm0.17$& $59.17\pm0.02$ \\ 
WDFS0238-36&02:38:24.969&-36:02:23.222&$24094\pm225$& $7.88\pm0.04$& $0.21\pm0.01$& $3.13\pm0.1$& $57.48\pm0.01$ \\ 
WDFS0248+33$^*$&02:48:54.965&33:45:48.244&$33646\pm483$& $7.29\pm0.06$& $0.34\pm0.01$& $3.36\pm0.08$& $58.32\pm0.02$ \\ 
WDFS0458-56$^*$&04:58:23.133&-56:37:33.434&$31128\pm218$& $7.79\pm0.04$& $0.04\pm0.01$& $3.1\pm0.17$& $57.9\pm0.01$ \\ 
WDFS0541-19&05:41:14.759&-19:30:38.896&$21144\pm195$& $7.88\pm0.03$& $0.1\pm0.01$& $3.25\pm0.16$& $57.54\pm0.01$ \\ 
WDFS0639-57$^*$&06:39:41.468&-57:12:31.164&$54352\pm1811$& $7.71\pm0.07$& $0.19\pm0.01$& $3.61\pm0.12$& $58.92\pm0.03$ \\ 
WDFS0727+32&07:27:52.752&32:14:16.046&$54754\pm3101$& $7.79\pm0.05$& $0.16\pm0.01$& $2.91\pm0.13$& $58.77\pm0.05$ \\ 
WDFS0815+07&08:15:08.782&07:31:45.775&$36439\pm1623$& $7.22\pm0.06$& $0.09\pm0.02$& $3.05\pm0.18$& $60.1\pm0.05$ \\ 
WDFS0956-38$^*$&09:56:57.009&-38:41:30.269&$19791\pm79$& $7.87\pm0.01$& $0.12\pm0.01$& $3.38\pm0.14$& $56.97\pm0.01$ \\ 
WDFS1024-00&10:24:30.912&-00:32:07.16&$36012\pm1350$& $7.79\pm0.13$& $0.23\pm0.02$& $3.06\pm0.15$& $59.09\pm0.05$ \\ 
WDFS1055-36$^*$&10:55:25.356&-36:12:14.731&$30682\pm199$& $7.93\pm0.03$& $0.15\pm0.01$& $3.31\pm0.11$& $58.0\pm0.01$ \\ 
WDFS1110-17$^*$&11:10:59.436&-17:09:54.308&$52672\pm1921$& $7.8\pm0.04$& $0.19\pm0.01$& $3.25\pm0.11$& $58.56\pm0.03$ \\ 
WDFS1111+39&11:11:27.313&39:56:28.105&$56759\pm3481$& $7.84\pm0.07$& $0.02\pm0.01$& $3.1\pm0.18$& $59.39\pm0.05$ \\ 
WDFS1206+02&12:06:50.41&02:01:42.138&$24243\pm272$& $7.96\pm0.02$& $0.08\pm0.01$& $3.12\pm0.16$& $58.22\pm0.01$ \\ 
WDFS1206-27$^*$&12:06:20.354&-27:29:40.639&$35884\pm361$& $7.8\pm0.04$& $0.15\pm0.01$& $3.35\pm0.11$& $56.77\pm0.01$ \\ 
WDFS1214+45$^*$&12:14:05.111&45:38:18.626&$36003\pm438$& $7.97\pm0.03$& $0.05\pm0.01$& $3.11\pm0.16$& $58.18\pm0.01$ \\ 
WDFS1302+10$^*$&13:02:34.422&10:12:38.717&$46154\pm887$& $7.87\pm0.02$& $0.1\pm0.01$& $2.87\pm0.11$& $57.71\pm0.02$ \\ 
WDFS1314-03&13:14:45.046&-03:14:15.685&$43685\pm2696$& $7.79\pm0.07$& $0.11\pm0.02$& $3.24\pm0.16$& $59.69\pm0.05$ \\ 
WDFS1434-28$^*$&14:34:59.528&-28:19:03.295&$20790\pm128$& $7.87\pm0.03$& $0.22\pm0.01$& $3.33\pm0.11$& $57.07\pm0.01$ \\ 
WDFS1514+00$^*$&15:14:21.277&00:47:52.380&$29853\pm74$& $7.88\pm0.01$& $0.18\pm0.0$& $3.45\pm0.08$& $55.6\pm0.01$ \\ 
WDFS1535-77$^*$&15:35:45.179&-77:24:44.832&$62316\pm1428$& $8.97\pm0.03$& $0.07\pm0.0$& $2.85\pm0.15$& $57.54\pm0.02$ \\ 
WDFS1557+55$^*$&15:57:45.38&55:46:09.361&$66877\pm1942$& $7.57\pm0.05$& $0.05\pm0.01$& $2.95\pm0.17$& $58.57\pm0.02$ \\ 
WDFS1638+00&16:38:00.352&00:47:17.739&$65645\pm3408$& $7.63\pm0.13$& $0.23\pm0.01$& $3.18\pm0.14$& $59.7\pm0.04$ \\ 
WDFS1814+78$^*$&18:14:24.078&78:54:03.084&$32019\pm144$& $7.91\pm0.02$& $0.05\pm0.0$& $3.06\pm0.14$& $56.75\pm0.01$ \\ 
WDFS1837-70$^*$&18:37:17.906&-70:02:52.513&$19975\pm91$& $7.89\pm0.02$& $0.14\pm0.01$& $3.02\pm0.12$& $56.88\pm0.01$ \\ 
WDFS1930-52$^*$&19:30:18.995&-52:03:46.55&$38498\pm632$& $7.79\pm0.04$& $0.17\pm0.01$& $3.33\pm0.11$& $57.86\pm0.02$ \\ 
WDFS2101-05&21:01:50.667&-05:45:51.159&$29696\pm358$& $7.86\pm0.05$& $0.17\pm0.01$& $3.22\pm0.13$& $58.54\pm0.02$ \\ 
WDFS2317-29$^*$&23:17:20.294&-29:03:21.647&$25063\pm224$& $7.82\pm0.03$& $0.08\pm0.01$& $2.83\pm0.16$& $57.97\pm0.01$ \\ 
WDFS2329+00&23:29:41.321&00:11:07.565&$21729\pm332$& $7.99\pm0.03$& $0.19\pm0.02$& $3.13\pm0.12$& $57.36\pm0.02$ \\ 
WDFS2351+37$^*$&23:51:44.274&37:55:42.569&$47766\pm1354$& $7.93\pm0.06$& $0.37\pm0.01$& $3.12\pm0.07$& $58.45\pm0.03$ \\ 
\hline
	\end{tabular}
\begin{flushleft}\small
\textbf{Notes:}  Coordinates are from \textit{Gaia} DR3 at epoch J2016.0 \citep{gaia_dr3}. Corresponding proper motions for each faint standard in our network can be found in Table 2 of \citet{a23}.
\end{flushleft}
\end{table*}

\begin{figure*}
\centering
	\includegraphics[width=1.8\columnwidth]{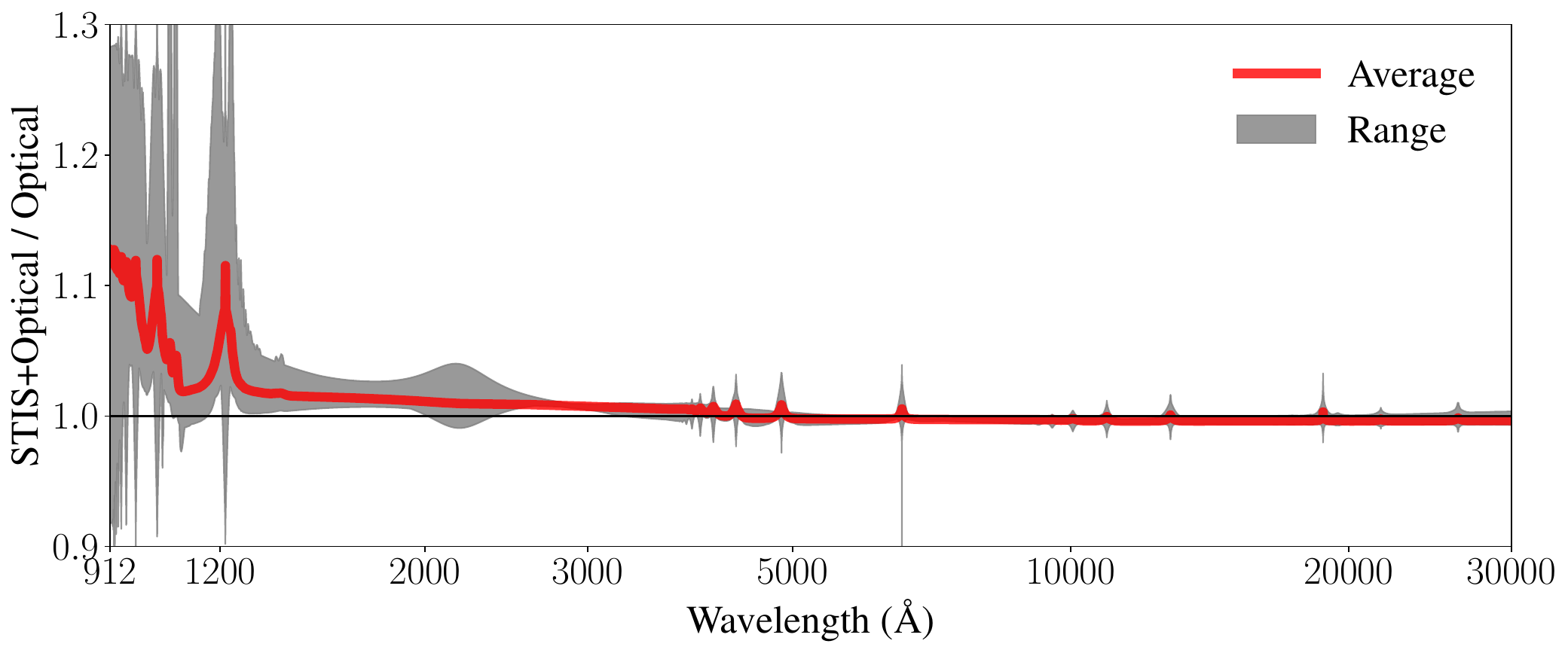}
    \caption{Ratio of SEDs inferred with and without using STIS UV spectra. Both SEDs were also inferred using optical spectra and HST/WFC3 photometry. The red line shows the average flux ratio as a function of wavelength, while the grey shows the range of ratios across the network of 35 stars. The sharp peak around 1200 \AA\space corresponds to the Lyman-$\alpha$ absorption feature that is on average 12\% stronger after the introduction of the STIS UV spectra.}
    \label{fig:stis_sed_comp}
\end{figure*}
\begin{figure*}
\centering
	\includegraphics[width=2.\columnwidth]{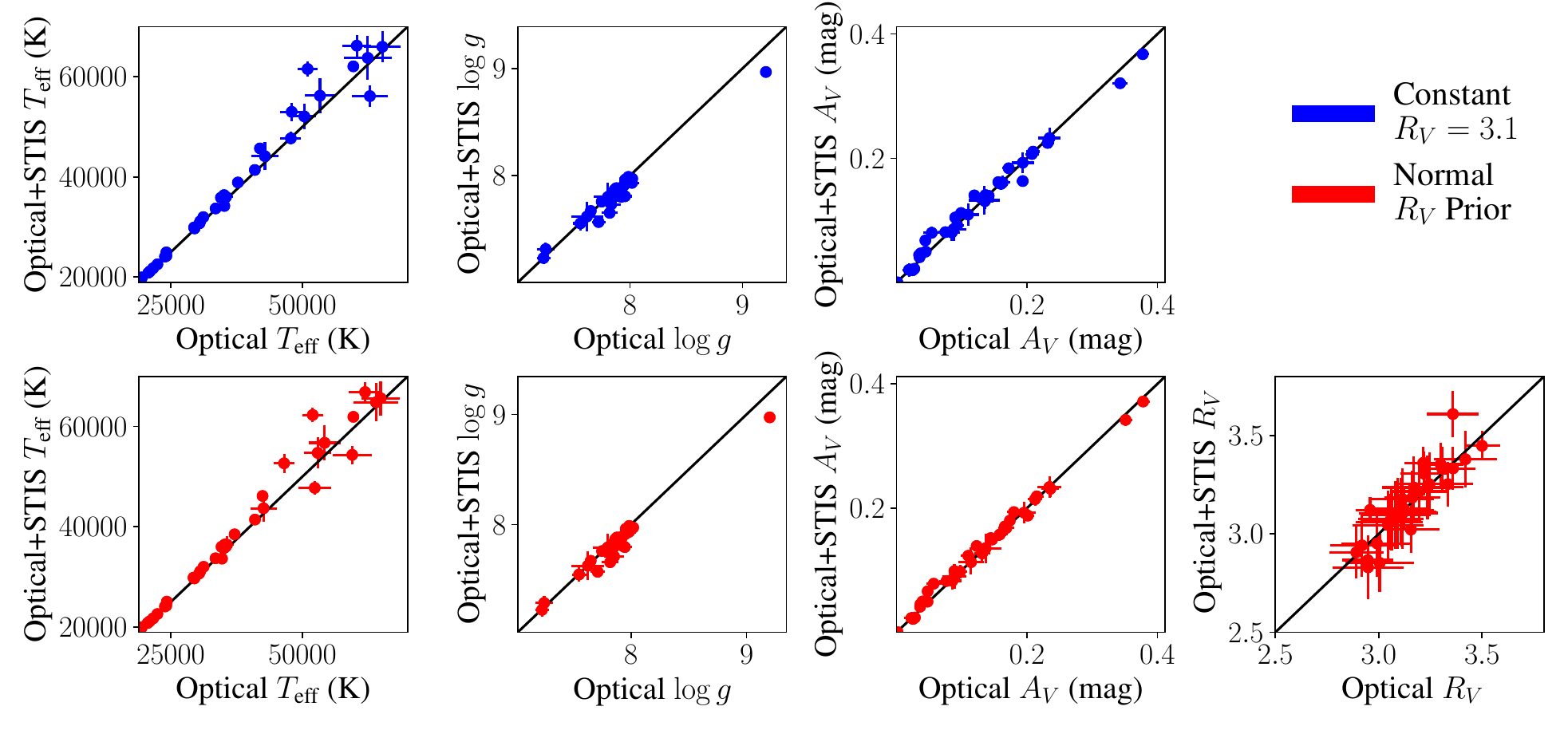}
    \caption{Comparing inferred parameters with and without STIS UV spectra. Inferences in top row use a constant $R_V=3.1$ for all objects, whilst the bottom row experiments jointly inferred the $R_V$ dust ratio for each object.}
    \label{fig:stis_param_comp}
\end{figure*}

\section{Results and Discussion}
\label{sec:results}
In this section we present our calibration results using our hierarchical Bayesian model. The main results presented are those that use all the optical ground-based spectra, the STIS UV space-based spectra for the 19 WDs that have coverage, and the six HST/WFC3 bands ranging the UV to the NIR. The published results are also those that infer $R^s_V$ for the WDs, rather than keeping it constant as in previous analyses. The prior used in the $R^s_V$ inference is a truncated normal prior $R^s_V\sim \mathcal{TN}(3.1,0.18,1.2,\infty)$ motivated by \cite{schlafly2016}, where $\sim$ denotes ``is sampled from''. In these main results we do not infer dust population parameters to avoid the effects of shrinkage, where inferences of object-level parameters are skewed towards the inferred population average. Our results are on the CALSPEC \cite{bohlin2020} magnitude system, tied at the F475W band, but may easily be converted to the \cite{bohlin2014b} system using the zeropoints presented in Table \ref{tab:zp_shift}. The SEDs for all 35 WDs in the network are defined across the entire range of the \cite{gordon2023} dust relation, spanning from 912 \AA\space to 32 $\mu$m, and are publicly accessible at: \url{https://zenodo.org/records/14339960}

\subsection{Inferred Parameters and SEDs}
\label{sec:seds}

The results of our inferred intrinsic and dust parameters for the 35 WDs are presented in Table \ref{tab:wd_params}. These five parameters per object are all that is necessary to forward model the calibrated SEDs. We illustrate the derived SEDs for the 32 faint standards, as well as the three CALSPEC primary standards, in Figure \ref{fig:all_seds}. 

Figure \ref{fig:stis_sed_comp} shows the average ratio between SEDs using the STIS UV spectra and those just using the optical spectra alone. Understandably, the largest change when adding this additional information was in the UV where the average change was around 12\% at the Lyman-$\alpha$ wavelength. The Lyman-$\alpha$ is the strongest absorption feature so it is logical that any deviation in inferred parameters will be most influential here. The grey change in flux around around 2175~\AA \space corresponds to differences in dust extinction at the  UV bump in the dust relation. To investigate further we plot the changes in parameters when introducing the STIS UV data in Figure \ref{fig:stis_param_comp}. We find the inferred $A_V$ dust extinction changes by 10\% on average after the introduction of the STIS UV spectra. Measuring such differences in extinction was part of the motivation behind obtaining UV spectra where dust effects are strongest. As expected, the estimated values of $R_V$ did not change significantly after the introduction of the STIS data. We would only expect $R_V$ values to change if we introduced high resolution NIR spectra so the model can see deviations of extinction across a range of wavelengths.  In Figure \ref{fig:stis_param_comp} we also see that the warmer WDs had greater changes in effective temperatures inferred after the introduction of the STIS data. This is an indication of a trade-off between having higher $A_V$ and $T_{\text{eff}}$ or having lower $A_V$ and $T_{\text{eff}}$. Further evidence of the trade-off is also illustrated in Figure \ref{fig:wd_corner}, where we see higher inferred temperature and higher inferred extinction for WDFS2317-29 after the introduction its STIS UV spectrum. The correlation can be explained by the increase in extinction decreasing the apparent strength of features, so the model needs to infer stronger temperatures to explain what is seen in the data. The introduction of the additional UV data had little effect on the inferred surface gravity for most of the WDs. The only exception was WDFS1535-77 that had a 4.04$\sigma$ decrease, from 9.20 to 8.97 when the new data was added, suggesting that the Lyman-$\alpha$ width is particularly important for modelling higher density DA WDs. This result may also be an indication that the source is influenced by magnetic fields that are broadening the Balmer absorption features via weak Zeeman splitting. 

We can also compare the new SEDs with those published in \cite{a23} in Figure \ref{fig:a23_sed_comp}. The differences in SEDs are due to numerous differences in methodology and modelling. The smooth decrease in average flux we see in the UV, including the bump at 2175 \AA, are due to the differences in dust relation when we transition from \cite{Fitzpatrick1999} to \cite{gordon2023} dust relations. In our work we jointly infer $R_V$ for each WD, whereas in previous works this was kept to a constant $R_V=3.1$. The sharp Balmer series flux increase of up to 7\% in the optical is likely due to the introduction of the STIS UV data influencing inferred $A_V$ and $T_{\text{eff}}$. In Figure \ref{fig:a23_param_comp_f99} we ignore the STIS data and choose the same modelling decisions as \cite{a23} finding we infer parameters that are largely consistent with the previous analysis. In Section \ref{sec:f99g23} we further explore the influence of using the \cite{gordon2023} dust relation over the \cite{Fitzpatrick1999} dust relation. Even after these modelling choices we would still expect to see different inferred SEDs between our work and the previous analysis, due to the changes in the updated theoretical DA WD template, the hierarchical nature of our model and the spectroscopic-informed priors in \cite{a23}.

In our main results we infer $R^s_V$ for each DA WD using a truncated normal prior with $\mu_{RV}=3.1$ and $\sigma_{RV}=0.18$, physically motivated by \cite{schlafly2016}. The distribution of inferred $R^s_V$ parameters do not deviate significantly from this prior using both \cite{gordon2023} and \cite{Fitzpatrick1999} dust relations. We find the average extinction $\tau$ does not deviate significantly above the prior value of 0.1 mag. We find WDFS2351+37 has a strong extinction of $A_V^s =0.372 \pm 0.009$ mag ($R_V^s=3.12\pm0.07$), which is indicative of anomalous reddening of this source. This finding is consistent with \cite{a23} and independent analysis by \cite{bohlin2025stis}. We also find strong extinction of $A^s_V=0.34\pm0.01$ mag ($R_V=3.36\pm0.08$) for WDFS0248+33, which is again consistent with \cite{a23}.

Since the \textit{DAmodel} statistical framework is hierarchical, we are also able to infer population parameters. As a demonstration of this we experiment by inferring a population $\mu_{RV}$, $\sigma_{RV}$ and $\tau$, as shown in Figure \ref{fig:rv_corner}. The inferred $\tau$ was $0.15 \pm 0.03$ mag which is consistent with the value used in our prior from our main analysis. The inferred population $\mu_{RV}$ was 3.42 $\pm$ 0.10 which is 3.2$\sigma$ away from the Milky Way average of $R_V=3.1$. The inferred standard deviation of the dust population was found to be $\sigma_{RV}$ = 0.39 $\pm$ 0.08, a factor of two higher than the literature value of $\sigma_{RV}$ = 0.18 from \cite{schlafly2016}. In the population $R_V$ results we are heavily influenced by selection effects, so our estimates should only be seen as a demonstration of what is possible with the framework. Furthermore, it is beneficial to ensure that the dust inferences for WD standards are independent and use physically motivated priors, so our published synthetic photometry and SEDs are those that are inferred using a $R_V$ prior with constant $\mu_{RV}$ and $\sigma_{RV}$ determined by the literature. In Section \ref{sec:future} we discuss the future possibility of using the framework on larger samples to gain more robust constraints on population statistics.

\subsection{Spectroscopic Spline Residuals}
\label{sec:splines}

An important part of our methodology involves inferring a cubic spline to account for residual differences between the observed spectroscopic data and the modelled spectral energy distribution. In Figure \ref{fig:spec_example} we show the model fit to WDFS1206-27 as an example. These splines may represent wavelength-dependent systematics in the ground or space-based spectrograph. Alternatively, the splines may be suggesting inadequacies in the modelling, where the theoretical SED model or the dust relation are unable to express the data. In \cite{n19} these residuals were modelled with a GP, but the functions themselves were found to be simple and smooth. In this work we instead chose to use a cubic spline with 10 equally spaced knots. We experimented with multiplying the splines in flux-space, but found this gave unreliable results, allowing absorption features to be artificially strengthened at different wavelengths, leading to differences in inferred parameters. Instead we settled on the adding the spline in flux-space, as this is still physically motivated by instrumentals effects and avoids degeneracies. In Figure \ref{fig:splines} we plot the relative strength of each inferred spline for the STIS UV and optical spectra.

The ground-based spectroscopic data may have been influenced by systematics introduced when calibrating to a set of standards stars. These standards themselves are only accurate to a few percent, whereas we require sub-percent accuracy. For this reason it was important to include extra flexibility to allow the model to deviate from the data. Our optical splines in the right panel of Figure \ref{fig:splines} show larger deviations from the model at the bluer (<4200 \AA) edges of the spectra. These differences were in agreement with what was found in \cite{n19} using GPs and are thought to be due to biased flux corrections at the edges of the spectrograph, where the throughput drops sharply.


Although STIS UV space-based spectra are less impacted by calibration systematics compared to optical spectra, it remains crucial to address other instrumental effects. In Figure \ref{fig:splines}, we see the strength of the inferred STIS UV splines decrease as a function of wavelength. The residuals are largely negative suggesting there may be an underestimation of flux in the STIS UV spectra. The STIS MAMA detectors dark current performance has degraded steadily in-flight and the dark current subtraction by the “On the Fly Reprocessing (OTFR)” pipeline may underestimate the dark level, leading to an additional background. As STIS uses two separate channels (FUV: 1150-1700 \AA\space and NUV: 1570-3180 \AA) and the final spectrum is reconstructed from both, the background level can exhibit a wavelength dependence as we see in the calibration splines, which increase in amplitude below 1700 \AA. Additionally, our faint DA standards require longer exposures than the bright standards used to calibrate STIS, and estimate the dark current rate, and  a small misestimation of the dark rate in the FUV MAMA would be sufficient to cause an additive flux excess in the final spectrum. However, we cannot uniquely identify the STIS dark current as the cause for the wavelength dependence seen in the STIS spectrum calibration splines, and in particular, sunlight reflected off the Earth or Moon can scatter onto the STIS detectors, and is dependent on the angle between each standard and the bright limb of the Earth/Moon, while scattered zodiacal light depend on helio-ecliptic coordinates\footnote{Information on STIS instrumental systematics: \url{https://www.stsci.edu/hst/instrumentation/stis/performance/background-noise}}. The intensity of scattered light is expected to drop as a function of wavelength, which is opposite to the trend we see in the STIS data. However, even if the intensity of scattered light drops in the UV, if the \emph{correction} for scattered light is underestimated in the UV, it could yield a chromatic effect similar to what is seen in our splines.
\begin{figure*}
\centering
	\includegraphics[width=2.\columnwidth]{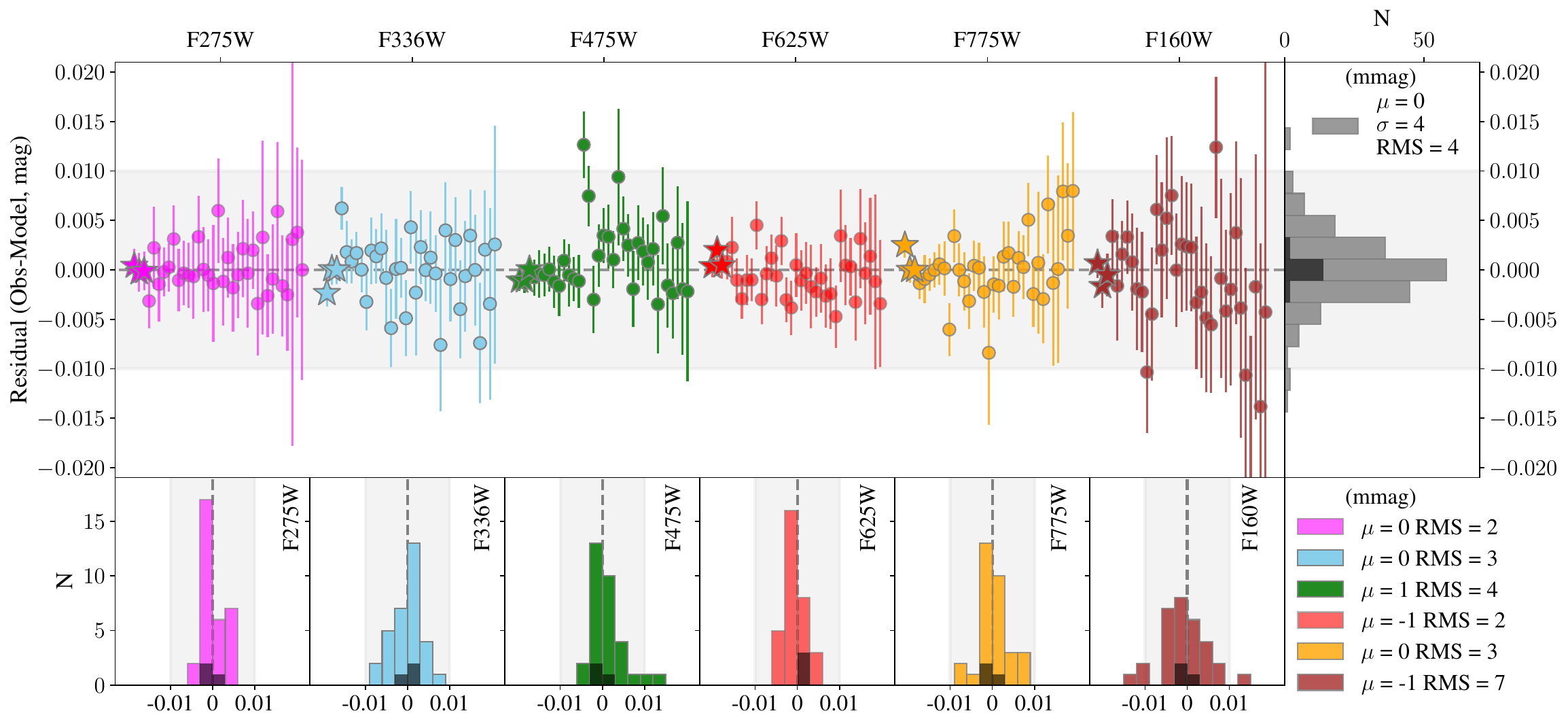}
    \caption{Final residual results achieved when using our model. Results show unweighted mean and RMS residuals for the 32 faint DA WDs, excluding the three bright CALSPEC standards. Scatter points are arranged in order of brightness, ending with the dimmest object. The results were found using the STIS UV spectra, optical spectra and the HST/WFC3 photometry. We use the \citet{gordon2023} dust relation and infer $R_V$ jointly per object.}
    \label{fig:resids}
\end{figure*}

The deviations between model and STIS UV spectra exhibited by the splines may also be hinting that the model needs to be improved at short wavelengths. The theoretical intrinsic DA WDs templates have recently been updated with improved UV modelling, but it is possible that more work needs to be done. Alternatively, there is a good chance that the differences may be due to the chosen dust relation. Dust relations like \cite{gordon2023} and \cite{Fitzpatrick1999} are averages of how extinction is observed to change as a function of $A_V$ and $R_V$. We find that the two strongest STIS splines belong to the object with the largest inferred $A^s_V=0.37$ and the object with the largest inferred $R^s_V=3.61$. If the dust relations are averages, it may be possible that these extreme regions of the $R_V-A_V$ space are poorly sampled leading to extinction curves that are at odds with our observed data. Besides these extremes, we find no obvious correlation between the dust parameters and inferred splines.

We are confident that the splines are not influencing the reliability of our published SEDs. Instead the splines are allowing us to utilise our data in a flexible manner that does not place too much emphasis on that relative calibration of each wavelength pixel. If the splines were influencing the inferred intrinsic parameters, we would see them trace the shapes of the Balmer series, either adding or removing signal. If the STIS splines were adding to extinction, we would expect them to trace the 2175 \AA\space bump. Conversely, we find our inferred splines to be smooth and to not exhibit traits that could be misinterpreted as additional signal coming from our model parameters. 
This conclusion is further supported by the observation that the inferred parameters in Figure \ref{fig:stis_param_comp}
do not exhibit the extreme change one would expect if they were artificially influenced by the STIS spline.

\begin{figure*}
\centering
	\includegraphics[width=1.6\columnwidth]{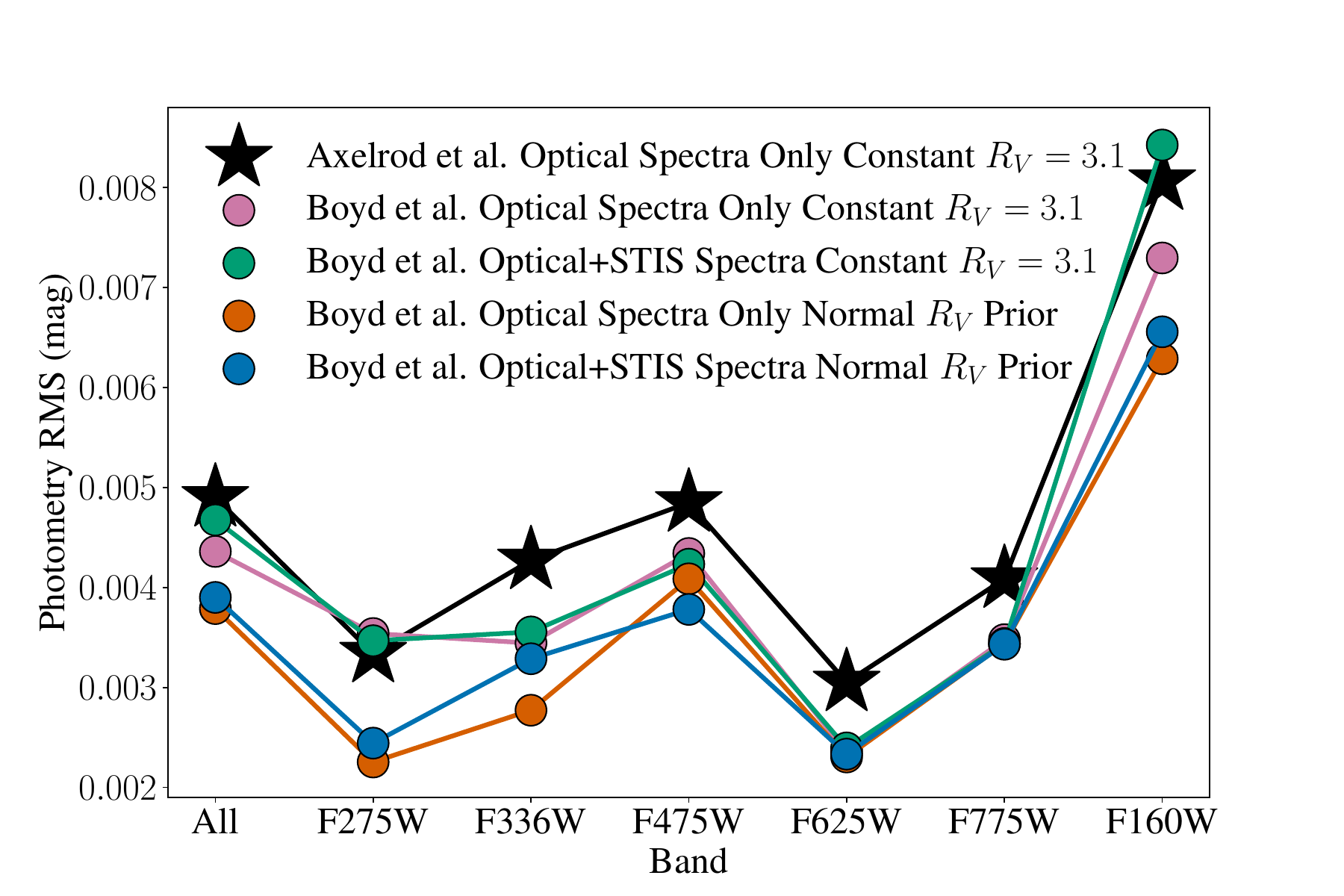}
    \caption{Comparison between photometric residuals of this work and \citet{a23}. The blue and green experiments utilised the STIS UV spectra for the 19 objects that had it, whilst the others experiments only used optical spectra. The blue and orange experiments used a truncated normal prior distribution to infer $R_V$ whilst the other experiments assumed it to be a constant 3.1. RMS residuals are unweighted including only the 32 faint DA WD stars and not the three CALSPEC standards.  }
    \label{fig:comp_a23}
\end{figure*}

\begin{table*}
	\centering
	\caption{Inferred HST/WFC3 photometric systematics obtained in our final results. This table is available in the online supplementary material.}

	\label{tab:zp}

	\begin{tabular}{lcccccc} 
 
		\hline
		Parameter & F275W&F336W&F475W&F625W&F775W&F160W \\
  \hline
$Z_X$ (mag)&23.978$\pm$0.002&24.544$\pm$0.001&25.572$\pm$0.001&25.407$\pm$0.001&24.732$\pm$0.001&25.789$\pm$0.002\\
$\Delta Z_X^{20}$ (mag)& -&-0.004$\pm$0.003&-0.02$\pm$0.003&-0.025$\pm$0.003&-0.02$\pm$0.004&-0.007$\pm$0.006\\
$\Delta Z_X^{22}$ (mag)&0.072$\pm$0.001&0.031$\pm$0.001&-0.007$\pm$0.001&-0.014$\pm$0.001&-0.028$\pm$0.001&0.002$\pm$0.001\\
$\sigma_{\text{int}}^X$ (mmag)&2.57$\pm$0.56&2.25$\pm$0.49&2.78$\pm$0.34&2.13$\pm$0.56&0.5$\pm$0.36&6.79$\pm$0.65\\
$\nu^X$&1.94$\pm$0.24&1.74$\pm$0.19&2.26$\pm$0.26&2.16$\pm$0.29&1.22$\pm$0.11&15.35$_{-6.57}^{26.04}$\\
$\alpha_X$ (mmag/mag)&-&-&-&-&-&-3.19$\pm$0.31\\
\hline
\end{tabular}
\begin{flushleft}\small
\textbf{Notes:} $Z_X$ represents the zeropoint magnitude at Cycle 25 (average MJD=57588.79 days) in a given band $X$. Zeropoints are calculated from our aperture radii (7.5 pixels for UVIS and 5 pixels for the IR) to infinity and are in AB mags. In Appendix \ref{sec:zp_shift} we compare inferred zeropoints to CALSPEC and relative shifts derived in \cite{a23}. $\Delta Z_X^{20}$ (average MJD=56452.94 days) and $\Delta Z_X^{22}$ (average MJD=57076.11 days) the change in zeropoint in each band in Cycle 20 and Cycle 22. We have no constraint on $\Delta Z_{\text{F275W}}^{20}$ as we only began F275W observations from Cycle 22. $\sigma_{\text{int}}^X$ and $\nu^X$ represent the student-T scale and dispersion parameters for each band. $\alpha_{\text{X}}$ represents CRNL parameter inferred only in the NIR F160W band. Estimates and uncertainties are the means and standard deviations of our chains, with the exception of $\nu^{\text{F160W}}$, that has a non-Gaussian posterior, so is presented as a median with the 16th and 84th percentiles. 
   
    \end{flushleft}
\end{table*}

\begin{table*}
	\centering
	\caption{HST/WFC3 synthetic photometry and residuals (labelled Synth. and Resid. respectively) from our final results. Units are in AB magnitudes and photometry for stars with an $*$ was inferred using STIS UV spectra as well as ground-based optical spectra. This table is available in the online supplementary material.}
	\label{tab:synth_phot}
	\begin{tabular}{lcccccccccccc} 
 
		\hline
		Object & \multicolumn{2}{c}{F275W}&\multicolumn{2}{c}{F336W}&\multicolumn{2}{c}{F475W}&\multicolumn{2}{c}{F625W}&\multicolumn{2}{c}{F775W}&\multicolumn{2}{c}{F160W} \\
        & Synth. & Resid. &Synth. & Resid. &Synth. & Resid. &Synth. & Resid. &Synth. & Resid. &Synth. & Resid.  \\
  \hline
G191B2B&10.481&4.1e-04&10.879&-0.002&11.488&-0.001&12.018&3.8e-04&12.433&0.002&13.866&0.001\\
GD153&12.194&-1.1e-04&12.554&2.0e-05&13.086&7.1e-05&13.586&4.5e-04&13.987&-4.0e-06&15.395&-0.001\\
GD71&11.979&-2.5e-04&12.319&6.7e-05&12.785&-0.001&13.264&0.002&13.657&-4.0e-06&15.048&-0.002\\
\hline
WDFS0103-00&18.18&0.006&18.517&-0.007&19.07&-0.002&19.556&0.001&19.942&0.008&21.318&-0.002\\
WDFS0122-30$^*$&17.661&0.001&17.981&-3.7e-04&18.447&0.001&18.916&-0.005&19.299&0.005&20.671&-0.004\\
WDFS0228-08&19.509&-3.0e-06&19.698&0.003&19.805&-0.002&20.158&-0.001&20.483&0.003&21.708&-0.014\\
WDFS0238-36&17.782&-0.001&17.959&-2.7e-05&18.082&0.001&18.427&4.9e-04&18.743&-0.002&19.954&0.002\\
WDFS0248+33$^*$&17.817&0.002&18.032&-0.008&18.354&0.003&18.736&-0.003&19.061&0.001&20.307&-0.005\\
WDFS0458-56$^*$&17.015&-0.001&17.335&0.002&17.743&-0.001&18.204&0.001&18.588&-0.003&19.956&0.008\\
WDFS0541-19&18.014&-0.003&18.199&0.003&18.262&0.002&18.615&-0.002&18.942&0.002&20.16&-0.003\\
WDFS0639-57$^*$&17.313&4.3e-05&17.639&9.3e-05&18.162&0.004&18.628&-0.001&19.002&-0.002&20.349&-0.006\\
WDFS0727+32&17.151&0.003&17.462&-0.006&17.979&0.001&18.445&-0.001&18.821&0.001&20.182&-0.002\\
WDFS0815+07&18.937&0.004&19.252&-0.003&19.705&-0.002&20.175&-0.003&20.557&0.008&21.923&-0.004\\
WDFS0956-38$^*$&17.691&-0.002&17.844&0.002&17.837&0.013&18.168&-3.9e-04&18.481&-9.0e-06&19.666&-0.01\\
WDFS1024-00&18.248&0.003&18.497&-9.0e-06&18.89&-0.002&19.303&0.003&19.651&-0.001&20.955&-0.004\\
WDFS1055-36$^*$&17.362&-0.001&17.64&2.0e-04&17.998&0.003&18.42&-0.004&18.779&-0.002&20.096&0.002\\
WDFS1110-17$^*$&17.032&-0.001&17.34&-0.001&17.847&0.007&18.304&-0.003&18.674&2.1e-04&20.017&0.003\\
WDFS1111+39&17.435&-0.001&17.817&-0.002&18.406&0.002&18.923&0.003&19.332&-0.002&20.753&0.004\\
WDFS1206+02&18.232&-0.002&18.468&0.003&18.661&-0.003&19.049&4.7e-04&19.395&0.001&20.667&-0.001\\
WDFS1206-27$^*$&15.729&-0.001&16.027&0.001&16.465&-4.3e-04&16.91&0.002&17.278&-0.001&18.619&-0.002\\
WDFS1214+45$^*$&16.931&-3.3e-04&17.266&0.001&17.749&-0.001&18.224&-0.001&18.614&4.2e-04&20.001&-3.0e-05\\
WDFS1302+10$^*$&16.178&2.7e-04&16.507&2.2e-05&17.025&-0.001&17.502&-0.001&17.888&0.001&19.269&0.001\\
WDFS1314-03&18.251&-0.003&18.578&0.002&19.084&0.003&19.557&-3.4e-04&19.94&9.0e-05&21.311&-0.022\\
WDFS1434-28$^*$&17.829&-3.4e-04&17.962&0.001&17.96&-0.003&18.271&0.003&18.569&-0.001&19.729&-0.004\\
WDFS1514+00$^*$&15.104&-0.003&15.369&0.006&15.697&-0.001&16.107&0.001&16.458&-0.001&17.754&0.003\\
WDFS1535-77$^*$&15.588&0.002&15.954&0.002&16.542&6.5e-05&17.042&-0.003&17.441&-1.0e-05&18.854&0.003\\
WDFS1557+55$^*$&16.487&0.003&16.865&-0.003&17.459&-0.002&17.974&0.005&18.379&-0.006&19.791&0.006\\
WDFS1638+00&18.003&0.003&18.305&-0.004&18.82&0.005&19.273&-0.003&19.638&0.007&20.969&-0.011\\
WDFS1814+78$^*$&15.782&-2.1e-04&16.105&0.002&16.532&-0.001&16.994&-0.001&17.378&-4.8e-04&18.751&0.002\\
WDFS1837-70$^*$&17.635&-0.001&17.774&0.004&17.76&-0.001&18.083&-0.003&18.392&0.003&19.573&-0.002\\
WDFS1930-52$^*$&16.722&-0.001&17.018&0.002&17.472&0.001&17.917&-0.001&18.284&1.6e-04&19.623&-0.002\\
WDFS2101-05&18.06&-0.001&18.32&-0.001&18.641&0.002&19.05&3.3e-04&19.402&-0.003&20.702&-0.002\\
WDFS2317-29$^*$&17.886&0.002&18.129&-0.001&18.34&-0.002&18.739&-0.002&19.089&2.8e-04&20.373&0.012\\
WDFS2329+00&17.937&-0.003&18.088&0.004&18.136&0.009&18.461&-0.002&18.768&-0.008&19.955&0.005\\
WDFS2351+37$^*$&17.434&0.006&17.652&-0.005&18.059&0.003&18.447&-3.4e-04&18.773&-0.001&20.035&0.002\\
\hline
\end{tabular}

\end{table*}

\subsection{HST/WFC3 Photometric Residuals and Systematics}
\label{sec:hst_sys}

In our work we achieve the lowest residuals yet when using our modelled SEDs to predict the observed data, illustrated in Figure \ref{fig:resids}. We use the 2000 samples (500 posterior samples from each of the four chains) of WD and dust parameters to forward model 2000 randomly sampled SEDs per object. To obtain the synthetic photometry we integrate the SED samples using Equation \eqref{eq:flux_int} over the six HST/WFC3 bandpasses described in \cite{c19,calamida2022}. We then define our synthetic photometry and uncertainty as the mean and standard deviation of the 2000 sampled magnitudes in each band for each object. We use the average of each inferred systematic (presented in Table \ref{tab:zp}) to correct for instrumental effects to derive the calibrated observed photometry. A detailed breakdown of the residual and synthetic photometry of each object is shown in Table \ref{tab:synth_phot}.

For the set of 32 faint standards, the average RMS from the UV to the NIR was 3.9 mmag, with an average bias of -0.06 mmag, using our synthetic SEDs and inferred systematics. Excluding the NIR band, the average RMS drops to 3.1 mmag, closer to the quantum efficiency (QE) limit. In Figure \ref{fig:comp_a23}, we compare these residuals (shown in blue) with those obtained by \cite{a23} (indicated by black stars), highlighting improvement across all bands. It is clear from the plot that these improvements are largely as a result of allowing $R^s_V$ for each band to be inferred, when in previous analyses this was kept constant. We find that switching from a constant $R_V=3.1$ to an free $R_V$ according to a truncated normal prior improves results significantly in the NIR F160W and UV F275W. This is predictable as $R_V$ is unlikely to be exactly 3.1 for the whole sample and differences in $R_V$ are more apparent at these wavelengths. From the plot it is also evident that adding STIS UV spectra data does not improve the RMS photometric residuals, however, we still consider these our main results as they are using the most information to determine the SEDs. 

We find small amounts of shrinkage when swapping from the normal $R_V$ prior with set mean and variance to the population prior where these are jointly inferred. Due to the additional parameters we find that the population $\mu_{RV}$ reduces the overall residual RMS to 3.6 mmag. We exclude these versions of the SEDs from our final results, as the selection effect heavily influences these population parameters, and we aim to avoid over-fitting the data.
Results of these tests and other experiments are presented in Table \ref{tab:tests}.

The procedure in \cite{n19} described a two-step process where first photometric measurements and offsets are inferred using a HBM, then these measurements are combined with spectra to serially infer WD parameters. As an investigation into our hierarchical results we serially infer WD parameters using a similar procedure to that described in \cite{n19} and find again there were no significant differences in inferred WD SED parameters. The differences in the results of the two approaches appeared in the inferred photometric systematics of the HST/WFC3 data. This difference is to be expected as our hierarchical method effectively changes the prior on $m^X_s$, that would have otherwise been flat in the two-step process, by introducing spectroscopic information. This effect was part of our motivation for using a fully hierarchical model, to allow for more precise inference of systematics by modelling all the data at once.

\begin{figure*}
\centering
	\includegraphics[width=2.\columnwidth]{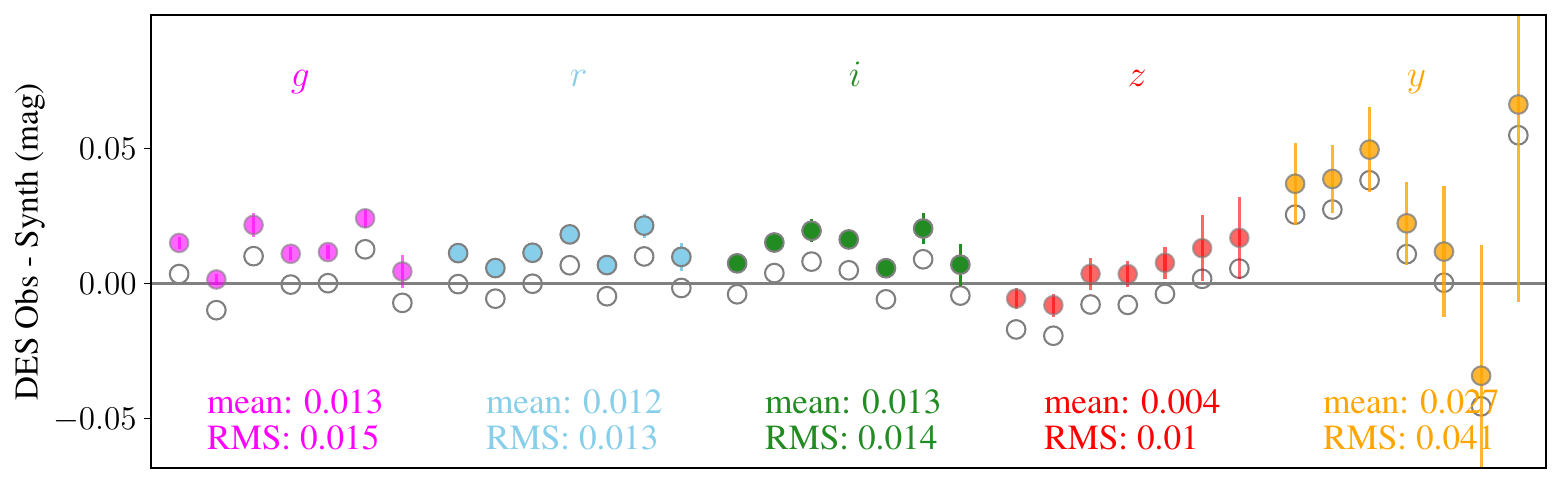}
    \caption{Photometric residuals when comparing our synthetic DES photometry with real DES observations \citep{des_dr2} for the faint DA WD standards that had coverage. Objects are arranged in order of brightness, ending with the dimmest object. We presented the unweighted mean and RMS residual for each band. Residuals are in AB magnitudes. Plotted in white are the residuals when using the \citet{bohlin2014b} CALSPEC system rather than the \citet{bohlin2020} system.}
    \label{fig:des_resid}
\end{figure*}
In Table \ref{tab:zp} we present our hierarchically inferred HST/WFC3 photometric systematics for each of the six bands. Zeropoints are calculated from our aperture radii (7.5 pixels for UVIS and 5 pixels for the IR) to infinity. The inferred $Z_X$ parameters correspond to UVIS1 and IR zeropoints at Cycle 25 (average MJD=57588.79 days) and are in 1\% agreement with those calculated in \cite{calamida2022zp} for the same aperture radius and epoch. In Appendix \ref{sec:zp_shift}, we compare zeropoint shifts in our work and the work of \cite{a23} with respect to the CALSPEC \cite{bohlin2014b} and \cite{bohlin2020} systems. We find good agreement with \cite{a23}, where we infer decreasing zeropoint shifts as function of wavelength ranging from 11 mmag in the UV to -15 mmag in the NIR for \cite{bohlin2020}. 

The inferred $\Delta Z_X^{20}$ and $\Delta Z_X^{22}$ are the deviations away from the Cycle 25 zeropoint in Cycle 20 (average MJD = 56452.94 days) and Cycle 22 (average MJD = 57076.11 days) respectively. We expect the $\Delta Z_X^{c}$ values to be negative if only due to a decrease of the WFC3 UVIS or IR detector sensitivity, as shown by
\cite{calamida2022zp} and \cite{marinelli2024}. The percentage in sensitivity decrease of each band is parametrised by a linear slope as a function of time. Since we gather our observations into discrete cycles, and the observations span a limited amount of time (< 5 years per star), it is challenging to compare with the official continuous slope. Despite this, we take averages of the percentage decrease evaluated at our observation times, finding that the $\Delta Z_X^{20}$ changes between Cycle 20 and Cycle 25 were within error of those calculated in \cite{calamida2022zp} and \cite{marinelli2024}. Our work infers a $\Delta Z_\text{F336W}^{20}$ consistent with zero which is also consistent with the literature. 

Cycle 20 observations were collected by using UVIS1 sub-arrays and are on the UVIS1 photometric system \citep{n16}, while Cycle 22 and Cycle 25 observations were instead collected by using UVIS2 sub-arrays. Cycle 22 images were then re-processed by omitting the default flux conversion performed by HST/WFC3 calibration pipeline (\textit{calwf3}), resulting in this photometry being on the UVIS2 photometric system \citep{c19}. Cycle 25 observations were not re-processed and the resulting photometry is on the UVIS1 photometric system \citep{a23}. These differences in photometric system do not influence our analysis, since the three primary DA WDs were observed in all cycles at multiple epochs allowing us to determine our own zeropoints and systematics.  

Our change in zeropoints for Cycle 22 are dominated by the effect of converting between the UVIS1 and UVIS2 photometric systems, rather than by changes in sensitivity, hence why some changes are positive. For instance, the difference between Cycle 25 and Cycle 22 zeropoints for the F275W is $\Delta Z_{\text{F275W}}^{22}\approx$ +0.07 mag, comparable to the QE difference between the UVIS1 and UVIS2 detectors for this filter, with the UVIS2 chip being more sensitive at bluer  wavelengths\footnote{HST/WFC3 CCD characteristics and performance:\\
\url{https://tinyurl.com/wfc3QE}}.
Note that this does not apply to the F160W filter, since there is only one WFC3 IR detector, so $\Delta Z_{\text{F160W}}^{20}$ and $\Delta Z_{\text{F160W}}^{22}$ are in this case mostly due to the change in sensitivity of WFC3-IR. We can compare the inferred changes zeropoints between Cycle 20 and Cycle 22 to those from \cite{n19} that used the same dataset. Comparisons find consistency in the difference between the $\Delta Z_X^{20}$ and $\Delta Z_X^{22}$ in our work and the shifts measured in \cite{n19}. 
 
 
An important systematic to consider is the CRNL in the F160W band. In our analysis, we infer the CRNL to be $\alpha_{\text{F160W}}$ = -3.19 $\pm$ 0.31 mmag/mag. This constraint is within $1\sigma$ of both the published CRNL value of -3.12 $\pm$ 0.32 mmag/mag from \cite{bohlin2019} and the combined result of -3.0 $\pm$ 0.24 mmag/mag from \cite{riess2019}. The result is also close to the post hoc result of  -2.6$\pm$0.52 mmag/mag from \cite{n19}. The constraint, however, is in disagreement with the result of -1.74 $\pm$ 0.32 mmag/mag from \cite{a23}, that used the same WD sample as this work. Interestingly, if we change our modelling decisions to match those of \cite{a23}, by using \cite{Fitzpatrick1999}, a constant $R_V=3.1$ and ignoring the STIS UV data, we obtain $\alpha_{\text{F160W}}$ = -2.12 $\pm$ 0.30 mmag/mag. This result  is a strong suggestion that the inferred value of the NIR CRNL is sensitive to dust modelling assumptions. In inference, our model is determining which effects are from natural sources such as dust and which are from instrumental systematics. If we change the definition of what is physically possible, such as fixing or changing dust relations, it will then impact the extent to which effects are inferred to be coming from systematics. The inferred CRNL's dependency on dust is also supported by correlations shown in Figure \ref{fig:sys_corner2}.

The inferred student-T dispersion $\sigma^X_{\text{int}}$ and degrees of freedom $\nu^X$ parameters were in close agreement with those inferred by \cite{n19} from the UV to the optical. The F160W parameter degree of freedom parameter $\nu^{\text{F160W}}$ was higher in our work, suggesting a normal distribution may have been sufficient, likely due to the better handling of the CRNL. The size of $\sigma^X_{\text{int}}$ inferred is comparable to our residual RMS in each band. This suggests that we are close to the limit of what can be achieved with this data and would need to address the instrumental systematics causing the inferred additional dispersion to achieve more accurate results.

\subsection{Photometric Residuals with other surveys}
\label{sec:other_res}
We demonstrate the use of our resulting SEDs to produce synthetic photometry for five common surveys. To obtain the synthetic photometry for each survey we repeat the same procedure as we describe at the beginning of Section \ref{sec:hst_sys}, but instead integrate over the bandpasses for each survey. We obtain the filter bandpasses from the Spanish Virtual Observatory (SVO) Filter Profile Service \citep{svo1,svo2}. For surveys like \textit{Gaia}, we must then do an extra step of converting our synthetic photometry from AB to Vega magnitude units. Only subsets of the 32 faint stars are observed by each survey, but it is still useful to look at the residual trends. The observations for each of the surveys are completely independent of the training of our model so we naturally expect to see larger scatter and biases than we would in the HST photometry. For detail on the observed photometric data obtained for each survey refer to \cite{n19} and \cite{a23}, where the same comparisons are made using their SEDs.

Figure \ref{fig:des_resid} shows the residual scatter for seven of our DA WDs in DES DR2 that were observed with DECam \citep{des_dr2}. We see residual bias in $g$, $r$ and $i$ bands between 12 and 13 mmag, which then reduces to 4 mmag in the $z$ band. We show corresponding results in Figure \ref{fig:sdss_resid} using SDSS \citep{holberg2006}, Figure \ref{fig:ps1_resid} using PS1 DR2 \citep{ps1_data}, Figure \ref{fig:gia_resid} using \textit{Gaia} DR3 \citep{gaia_dr3} and Figure \ref{fig:decals_resid} using DECaLS \citep{decals} data. Residual bias with other surveys is highly dependent on the magnitude systems used in calibration of both our work and the surveys themselves. 
In our method we set the primary standard magnitudes as constants in one band according to \cite{bohlin2020} CALSPEC. The other surveys that we calculate residuals with have been calibrated with respect to the CALSPEC \cite{bohlin2014b} system or older. In Figure \ref{fig:des_resid} and the Figures in Appendix \ref{sec:other_res} we also show the residuals if we were to use the CALSPEC \cite{bohlin2014b} system. These results are comparable to those in \cite{a23}. This change to the older system dramatically reduces the size residuals, particularly with \textit{Gaia} in Figure \ref{fig:gia_resid}, where we have an average residual of -1 mmag in the $G$ band when using the \cite{bohlin2014b} system and the same scaled definition of Vega \citep{riello2021}. Despite getting better agreement with other surveys when using the older CALSPEC system, we maintain our decision to publish the network with respect to the more accurate \cite{bohlin2020} CALSPEC system and encourage other surveys to adopt the same for their absolute calibration.

These comparisons between our synthetic photometry and observed photometry from other surveys are necessary for cross-calibration when combining datasets. Our all-sky faint standards will allow various surveys to use a single self-consistent flux scale. In Section \ref{sec:future} we discuss efforts that are already using our inferred SEDs in the cross-calibration of datasets.

\begin{figure}
\centering
	\includegraphics[width=1.\columnwidth]{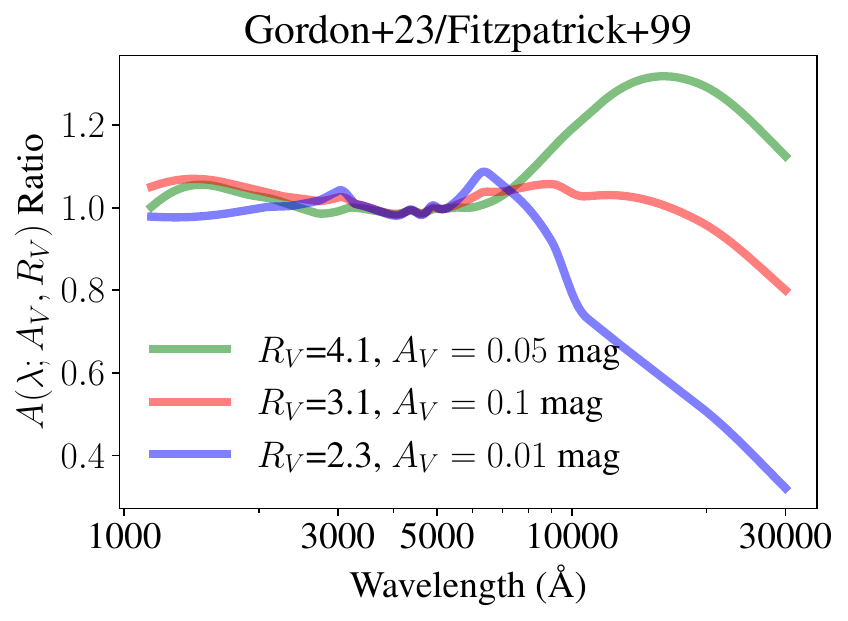}
    \vspace{-0.5cm}
    \caption{The ratio of \citet{gordon2023} dust relation with \citet{Fitzpatrick1999} dust relation, evaluated at different $A_V$ and $R_V$ values. We see moderate discrepancies in the UV and strong discrepancies in the NIR.}
    \label{fig:f99_g23}
\end{figure}
\begin{figure*}
\centering
	\includegraphics[width=2.\columnwidth]{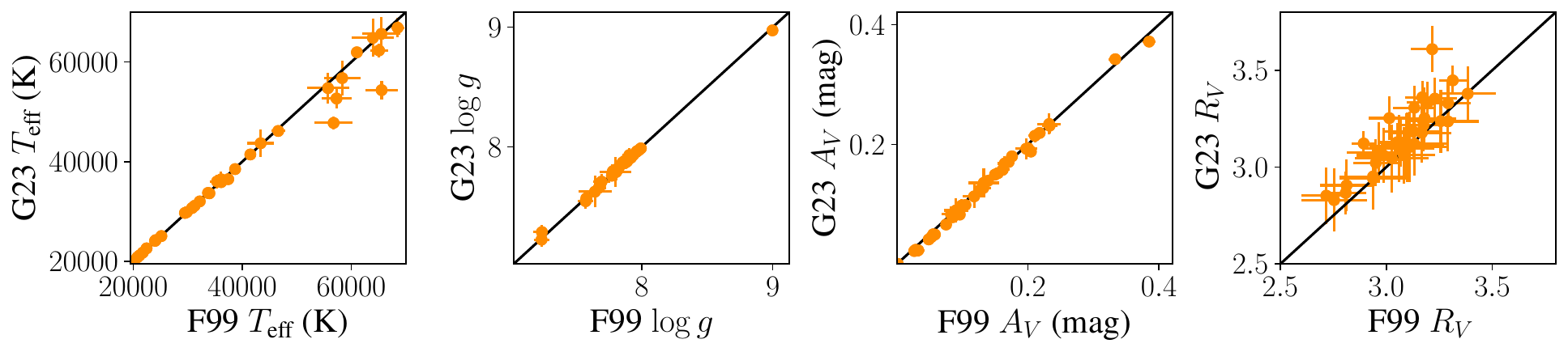}
    \caption{Comparing inferred parameters when using \citet{gordon2023} dust relation and \citet{Fitzpatrick1999} dust relation.}
    \label{fig:f99_param_comp}
\end{figure*}
\subsection{Using G23 Over F99 Dust Relation}
\label{sec:f99g23}
An important update to our modelling of the DA WDs was the adoption of the \cite{gordon2023} dust relation  over  the \cite{Fitzpatrick1999} relation. As illustrated in Figure \ref{fig:f99_g23}, the main differences between the relations appear in the UV and NIR wavelengths, which are critical regimes for robust inference of $A_V$ and $R_V$. It is important that our models extrapolate well to wavelengths  where we do not have data coverage, which is heavily reliant on the accuracy of the dust relation.

In Figure \ref{fig:f99_param_comp} we compare the inferred parameters when using both the \cite{Fitzpatrick1999} and \cite{gordon2023} dust relations with a truncated normal $R_V$ prior whilst utilising the STIS UV spectra. There is good agreement for $\log g^s$ surface gravity inference when using the different dust relations. Switching to \cite{gordon2023} results in a 7\% decrease in $A_V$, causing the warmer DA WDs to have lower inferred effective temperatures. The two WDs that deviate the most had STIS UV coverage, suggesting there was less extinction at these wavelengths, decreasing the inferred intrinsic strength of the Lyman-$\alpha$ absorption features. Switching to the updated relation results in $R_V^s$ inferences that differ on average 2\% greater, hinting a slight change in definition, although the uncertainties on this parameter are still large due to the lack of spectroscopic coverage in the NIR. The $R^s_V$ of WDFS0639-57 increases significantly from 3.22 $\pm$ 0.13 to 3.61 $\pm$
0.10 when switching form \cite{Fitzpatrick1999} to \cite{gordon2023}, again emphasising a change in definition of the dust relations at parameter extremes.
During our population analysis we see large differences in the inferred $\mu_{R_{V}}$, increasing from 3.01 $\pm $ 0.10 to 3.42 $\pm$ 0.10 when switching to the newer relation, however, as previously stated, these results are still influenced heavily by selection effects and small sample sizes.

In Table \ref{tab:tests} we can study the effect switching dust relation has on HST/WFC3 residuals. We see using \cite{gordon2023} dust relation results in a lower average residual bias and RMS scatter across the six bands. The largest improvement was in the F275W UV band where both bias and scatter dropped by roughly 1 mmag. Interestingly, results were marginally worse in the F475W and F625W bands  when using \cite{gordon2023}, although these results are already close to QE.

Dust relations are averages so will never completely agree with our observed data. In Section \ref{sec:future} we discuss the possibility of extending our model to simultaneously infer its own dust relation.

\section{Future Work}
\label{sec:future}

It is a priority to obtain NIR spectra for our faint DA WD standards in the near future by using an instrument such as NIRSpec on JWST \citep{nirspec}. Note that a sub-sample of these stars are already included in the official JWST calibration program (see proposal 6605, PI Gordon).
Adding spectroscopic coverage at IR wavelenghts would allow for more robust inferences of $R_V$ and further reduce photometric residuals in the F160W band. Modelling the NIR spectra would be a trivial extension to the model, as it was for the STIS UV spectra, where we simply evaluate the model at a different set of wavelengths and include additional cubic splines. We are also eager to obtain further STIS UV coverage for the remaining 13 DA WDs that do not have it, indicated in Table \ref{tab:wd_params} and Table \ref{tab:synth_phot} by the WDs without an asterisk. Obtaining this panchromatic spectroscopic coverage will improve the network's ability to be NIR standards for JWST itself, as well as Roman and the LSST $y/z$ bands.

The scalable implementation of the model allows for the possibility of expanding the network to more objects. Natural targets to join the network would be the 11 DA WDs observed by \cite{rubin2022prop} using HST/STIS. The 17 cooler DA WDs added to CALSPEC by \cite{elms2024} could also be a extension to the network providing we update our theoretical intrinsic DA WD templates at these temperatures. Hierarchically modelling this extended sample of DA WDs would further improve internal HST/WFC3 systematic mitigation, whilst having higher statistics would allow future surveys to improve their flux calibration.

Aside from the purpose of calibration, our novel statistical framework opens doors to population analysis on larger samples of DA WDs. There are nearly 2000 DESI DA WDs \citep{manser2024} and a further 500 DA WDs in the \textit{Gaia} 40 pc sample \citep{obrien2024}, with both sets having spectroscopic and photometric coverage. Analysing these volume-limited subsets with our framework could provide more meaningful statistics on population dust parameters. Alternatively, we could take the same approach as \cite{fm1990} by adding additional parameters to the extinction function to infer our own dust relation. We could also extend the framework to provide population inferences on the surface gravity and effective temperature parameters. If we introduce \textit{Gaia} parallaxes to our spectrophotometric framework,
we could use the inferred surface gravity to estimate mass evolution of DA white dwarfs \citep{shijing2017,shijing2018}. These ideas would not have been computationally feasible in the past, but are now within reach thanks to our GPU accelerated implementation of \textit{DAmodel}. 

An important science case that motivates the need for our all-sky accurate flux standards is SN Ia cosmology. Most recent efforts to constrain our Universe use compilations of SNe from a variety of surveys \citep{scolnic2022,brout2022,rubin2023}. An essential part of the analysis is the cross-calibration of the surveys so they are on a consistent flux scale \citep{scolnic2015}. Popovic et al. (2025, in prep.) are already using our inferred SEDs in their \textit{Dovekie} cross-calibration codebase. Their work will combine the present day's largest SNe Ia samples from the Zwicky Transient Facility \citep{ztf2024}, Dark Energy Survey \citep{des2024} and more, to further constrain the nature of our Universe. 

\section{Conclusions}
\label{sec:concl}
We have developed a novel hierarchical Bayesian statistical framework for the spectrophotometric calibration of 32 faint DA white dwarfs ($16.5 < V <19.5$) alongside three CALSPEC primary standards, from the UV to the NIR. We simultaneously utilise data from a variety of sources, including ground-based optical spectra, new space-based UV HST/STIS spectra and six HST/WFC3 photometric bands. In our framework we develop a flexible method of inferring a cubic spline surface to account for residual spectrograph instrumental systematics and inadequacies of the model. In parallel we jointly infer complex photometric systematics relating to time-varying sensitivity and count-rate nonlinearity. The model uses the most up to date data, DA WD templates and dust relations to allow for the most accurate modelling of the network yet; yielding an average HST/WFC3 RMS of less than 4 mmag from the UV to the NIR.

Including space-based HST/STIS UV spectra results in $A_V$ inferences that differ by an average of 10\% compared to those derived without the UV data. We find that the additional UV coverage has little impact on $R_V$ inference, indicating that NIR spectra are also necessary for tighter constraints. Further improvements in modelling were achieved by adopting the dust relation from \cite{gordon2023}, which resulted in smaller residuals compared to the \cite{Fitzpatrick1999} dust relation. By design, the hierarchical structure of our model does not affect individual DA WD parameter inferences, but instead enhances the precision of shared instrumental systematics estimation. When modelling HST/WFC3 photometry, we infer a NIR F160W count-rate nonlinearity of -3.19 $\pm$ 0.31 mmag/mag, in agreement with previous studies.

The model has seen vast performance improvements since \cite{n19} thanks to the combination of using Hamiltonian Monte Carlo, GPU acceleration and cubic splines rather than Gaussian Processes. This has resulted in convergence time going from a day per object to half an hour for the full sample. These speed-ups provide new opportunities to expand the model to include more WDs and further increase spectroscopic coverage. Additionally, we can apply the framework to larger samples for population-level parameter inference. The methods developed in this work will also inspire spectroscopic training of the next iteration of \textit{BayeSN}, a hierarchical Bayesian model for SN Ia modelling \citep{mandel2009,mandel2011,mandel2020,thorp2021,grayling2024}.

The faint all-sky DA white dwarf network will be imperative for the future of SN Ia cosmology. Work is already underway using the SEDs inferred from this work to combine today's largest samples of SN Ia. With the addition of near-infrared spectra, the network will become even more useful in the calibration and synergies of current and next-generation surveys, including the James Webb Space Telescope, Vera Rubin Observatory’s Legacy Survey of Space and Time and the Nancy Grace Roman Telescope.

\section*{Acknowledgements}

BMB is supported by the Cambridge Centre for Doctoral
Training in Data-Intensive Science funded by the UK Science and Technology Facilities Council (STFC) and two G-Research early career researchers grants used for equipment and travel. BMB also acknowledges travel support provided by Christ's College Cambridge's Clayton Fund, LSST DESC supported by DOE funds administered by SLAC and STFC for UK participation in LSST through grant ST/S006206/1. BMB, AB, ML and GN were supported on this project through the HST Guest Observer Program (HST-GO-16764, PI G. Narayan). GN also gratefully acknowledges NSF support from AST-2206195, and a CAREER grant AST-2239364, supported in-part by funding from Charles Simonyi, and OAC-2311355, DOE support through the Department of Physics at the University of Illinois, Urbana-Champaign (\#13771275), and HST-GO-17128 (PI R. Foley). Supernova and astrostatistics research at Cambridge University is supported by the European Union’s Horizon 2020 research and innovation programme under European Research Council Grant Agreement No 101002652 (PI K. Mandel) and Marie Skłodowska-Curie Grant Agreement No 873089.  CWS acknowledges support from Harvard University and the DOE cosmic frontier program through award DE-SC0007881. EO was also partially supported by NSF grants  AST-1815767 and AST-1313006. We acknowledge support from STSCI/HST: HST-GO-12967, HST-GO-13711, HST-GO-15113. 

Spectroscopic observations reported here were obtained at the MMT Observatory, a joint facility of the University of Arizona and the Smithsonian Institution. Spectra were also obtained at SOAR and at Gemini based in part on observations obtained at the Southern Astrophysical Research (SOAR) Telescope, which is a joint project of the Ministério da Ciência, Tecnologia e Inovações do Brasil (MCTI/LNA), the US National Science Foundation’s NOIRLab, the University of North Carolina at Chapel Hill (UNC), and Michigan State University (MSU). Observations were also obtained at the international Gemini Observatory, a program of NSF’s NOIRLab, which is managed by the Association of Universities for Research in Astronomy (AURA) under a cooperative agreement with the National Science Foundation on behalf of the Gemini Observatory partnership: the National Science Foundation (United States), National Research Council (Canada), Agencia Nacional de Investigaci\'{o}n y Desarrollo (Chile), Ministerio de Ciencia, Tecnolog\'{i}a e Innovaci\'{o}n (Argentina), Minist\'{e}rio da Ci\^{e}ncia, Tecnologia, Inova\c{c}\~{o}es e Comunica\c{c}\~{o}es (Brazil), and Korea Astronomy and Space Science Institute (Republic of Korea). This work makes use of observations from the Las Cumbres Observatory global telescope network. This research has made use of the SVO Filter Profile Service "Carlos Rodrigo", funded by MCIN/AEI/10.13039/501100011033/ through grant PID2023-146210NB-I00.

AS dedicates his efforts towards this project to the memory of his late advisor Dr. J.B. Oke, whose measurements of the absolute flux distribution of Vega pioneered the application of spectrophotometry in astronomy.

\section*{Data Availability}

The implementation of \textit{DAmodel} v1.0 is publically available at: \url{https://github.com/benboyd97/DAmodel}. All data used, coordinates and our published v1.0 SEDs can be found at: \url{https://zenodo.org/records/14339960}. The theoretical DA WD spectra \citep{hubeny2021} used to create the underlying grid can be found at: \url{www.as.arizona.edu/~hubeny/pub/DAgrid24.tar.gz}. The SEDs used to tie our work to the CALSPEC system \citep{bohlin2020} may be found at: \url{https://archive.stsci.edu/hlsps/reference-atlases/cdbs/current_calspec/}. The HST/WFC3 UVIS2 and IR filters \citep{calamida2022zp} used in this work are available at: \url{https://www.stsci.edu/hst/instrumentation/wfc3/performance/throughputs}. Data from other surveys used to calculate residuals is published in \cite{a23}.



\bibliographystyle{mnras}
\bibliography{paper} 



\newpage
\onecolumn
\newpage

\appendix
\section{Published SEDs}
\begin{figure*}
\centering
	\includegraphics[width=0.9\columnwidth]{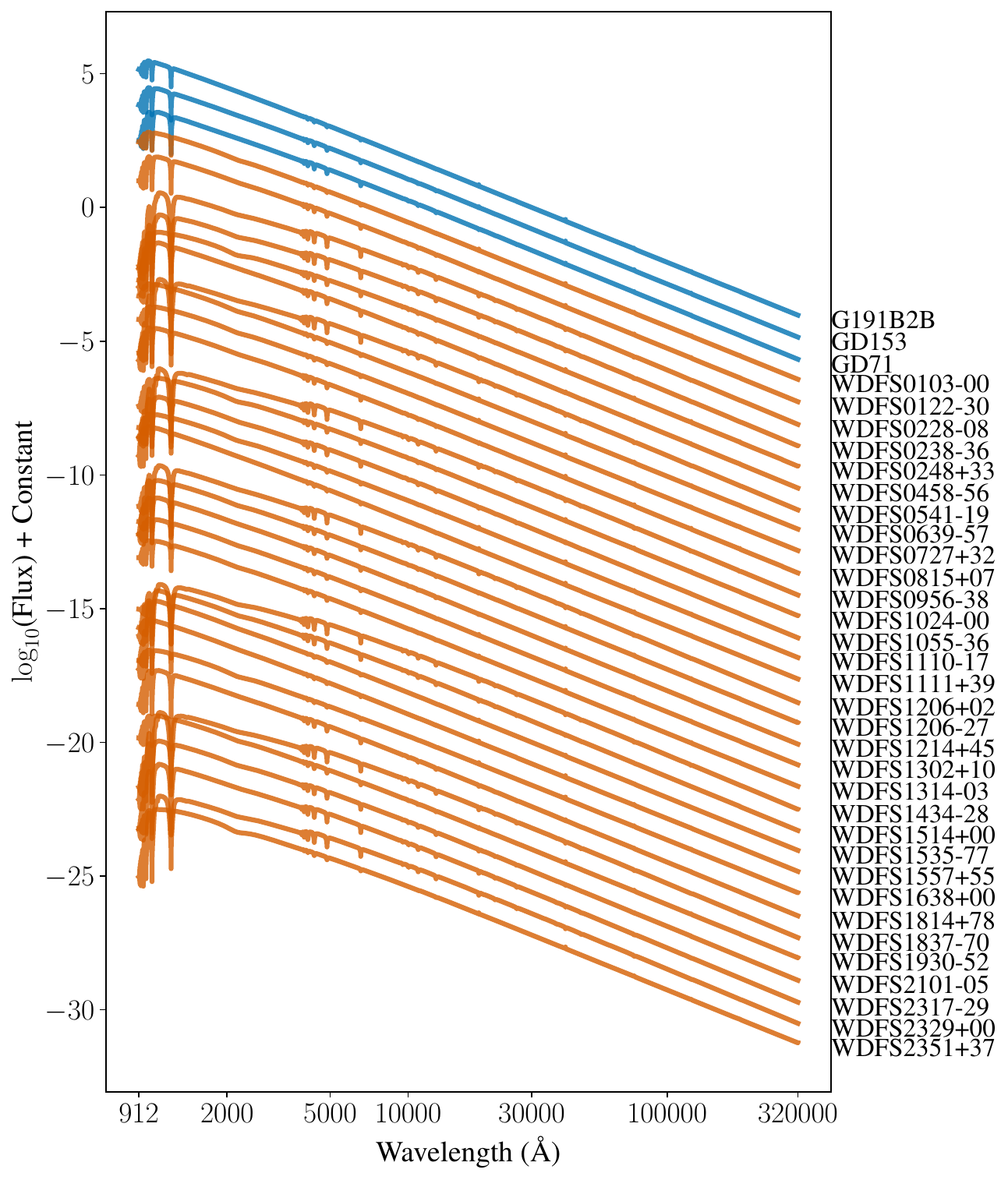}
    \caption{Illustration of our 35 inferred SEDs for the DA WD network. The top three SEDs in blue are the three bright CALSPEC standards, while the rest in orange belong to the faint standards. }
    \label{fig:all_seds}
\end{figure*}

\begin{figure*}
\centering
	\includegraphics[width=1\columnwidth]{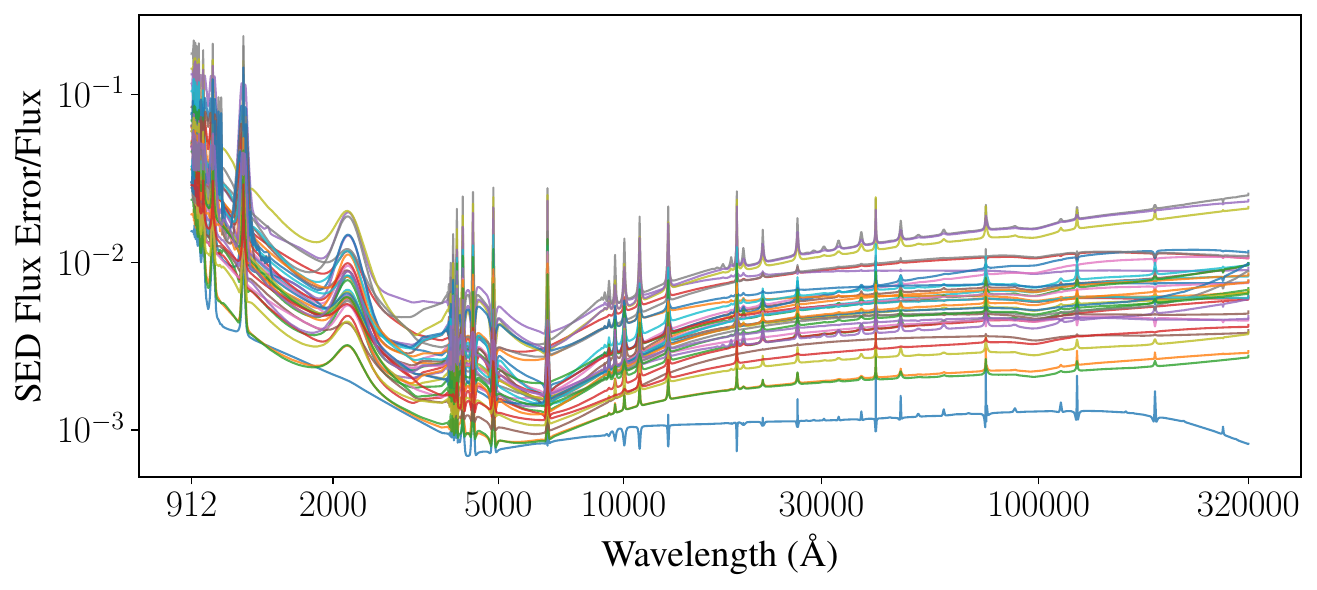}
    \caption{Inferred SED flux error fraction as a function of wavelength for the DA WDs in our network. These errors are estimated by sampling the inferred WD parameters to generate random realisations of the SEDs, with the standard deviation at each pixel representing the uncertainty. We use the average of the sampled SEDs to determine the flux at each pixel.}
    \label{fig:seds_err}
\end{figure*}

In Figure \ref{fig:all_seds} we illustrate the SEDs for each WD in our network that we publish alongside our work. Each published WD SED is derived by averaging 2000 randomly drawn SEDs forward-modelled from 2000 posterior samples. With these SED draws we may also find the associated uncertainty at each wavelength found by taking the standard deviation at each pixel, shown in Figure \ref{fig:seds_err}. Our published SEDs and associated errors, as well as coordinates, can be found at: \url{https://zenodo.org/records/14339960}.

\begin{table*}
	\centering
	\caption{Resources released with this work, along with those utilised in its methodology.}
	\label{tab:links}
    
	\begin{tabular}{p{8cm}p{8cm}} 
    \hline
Resource & Location \\
\hline
Full coordinates of the 35 DA WDs & \url{https://zenodo.org/records/14939532/files/wdfs_coords.txt}\\
35 DA WD SEDs and corresponding errors derived from v1.0 of this work, along with all spectroscopic and photometric data utilised& \url{https://zenodo.org/records/14339960}\\
Tables 1-3 and synthetic photometry for other surveys & \url{https://zenodo.org/records/16877348/files/paper_tables.zip}\\
Implementation of \textit{DAmodel} v1.0 used in this work &\url{https://github.com/benboyd97/DAmodel}\\
Theoretical DA WD spectra computed with \texttt{Tlusty} v208 \citep{hubeny2021}  &\url{www.as.arizona.edu/~hubeny/pub/DAgrid24.tar.gz}\\
SEDs used to tie the network to the CALSPEC system: g191b2b\_mod\_012.fits,  gd153\_mod\_012.fits and  gd71\_mod\_012.fits \citep{bohlin2020} & \url{https://archive.stsci.edu/hlsps/reference-atlases/cdbs/current_calspec/}\\
HST/WFC3 UVIS2 and IR filters \citep{calamida2022zp}
& \url{https://www.stsci.edu/hst/instrumentation/wfc3/performance/throughputs}\\
Data from other surveys used to calculate residuals & Published in \cite{a23} \\
\hline
\end{tabular}
\end{table*}
\clearpage
\section{Comparisons With Axelrod et al. (2023)}

\begin{figure*}
\centering
	\includegraphics[width=1.\columnwidth]{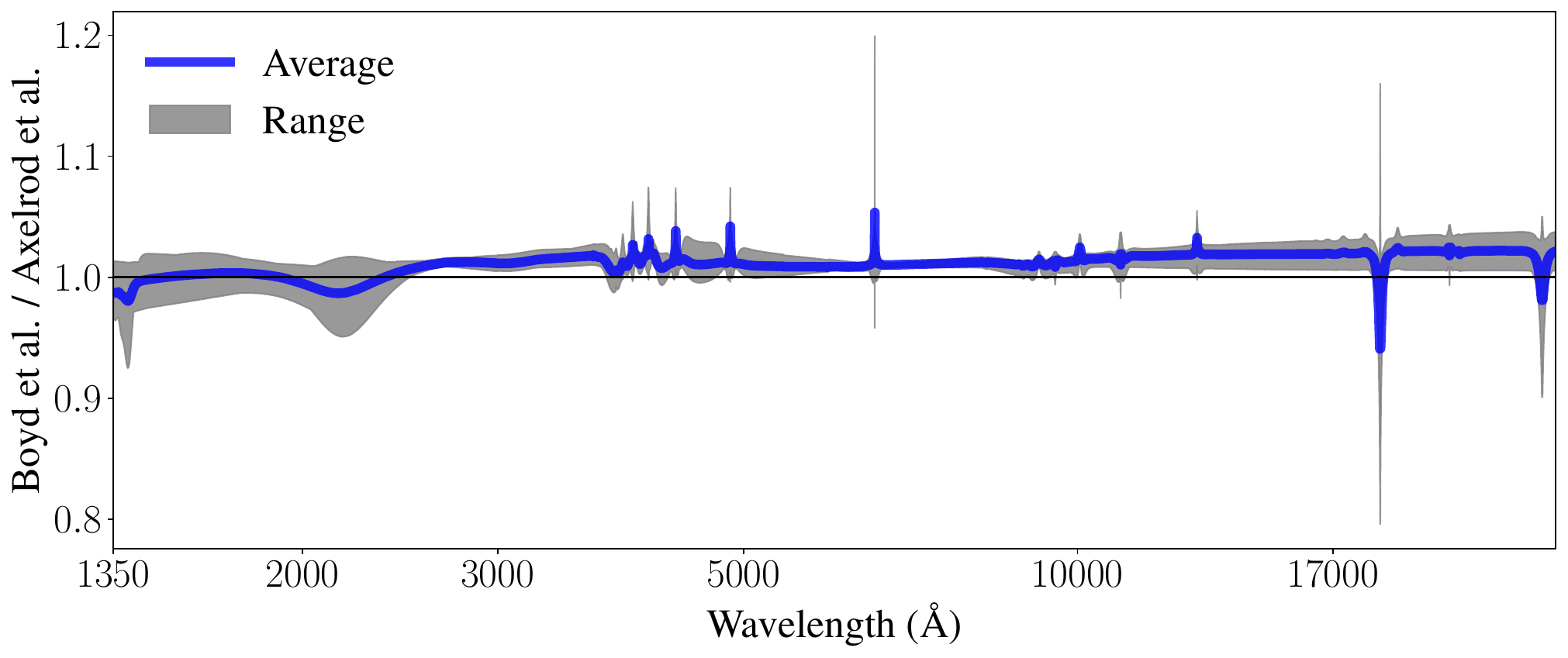}
    \caption{Ratios of SEDs inferred from this work and those from \citet{a23}. Differences come from using different theoretical SED templates, dust relations, $R_V$ assumptions, spectroscopic data and methodology. The blue line indicates the average ratio as a function of wavelength, while the grey shaded region shows the range of ratios across the 32 faint WDs.}
    \label{fig:a23_sed_comp}
\end{figure*}

\cite{a23} previously analysed the same network of stars. We plot the average ratio between our inferred SEDs and those from \cite{a23} at wavelengths with mutual coverage in Figure \ref{fig:a23_sed_comp}. The largest differences are observed in the UV, likely driven by our inference of the $R_V$ parameter, the adoption of the updated dust relation and theoretical template, and the inclusion of HST/STIS UV spectra. Spikes in the optical are due to differences in inferred effective temperature, resulting in differing absorption feature strength. In Figure \ref{fig:a23_phot_comp} we show the difference in synthetic HST photometry between our work and \cite{a23}. The residuals are mostly driven by the adoption of the updated \cite{bohlin2020} CALSPEC system, whilst mmag-level deviations are caused by differences in modelling choices and methodology.

\begin{figure*}
\centering
	\includegraphics[width=1.\columnwidth]{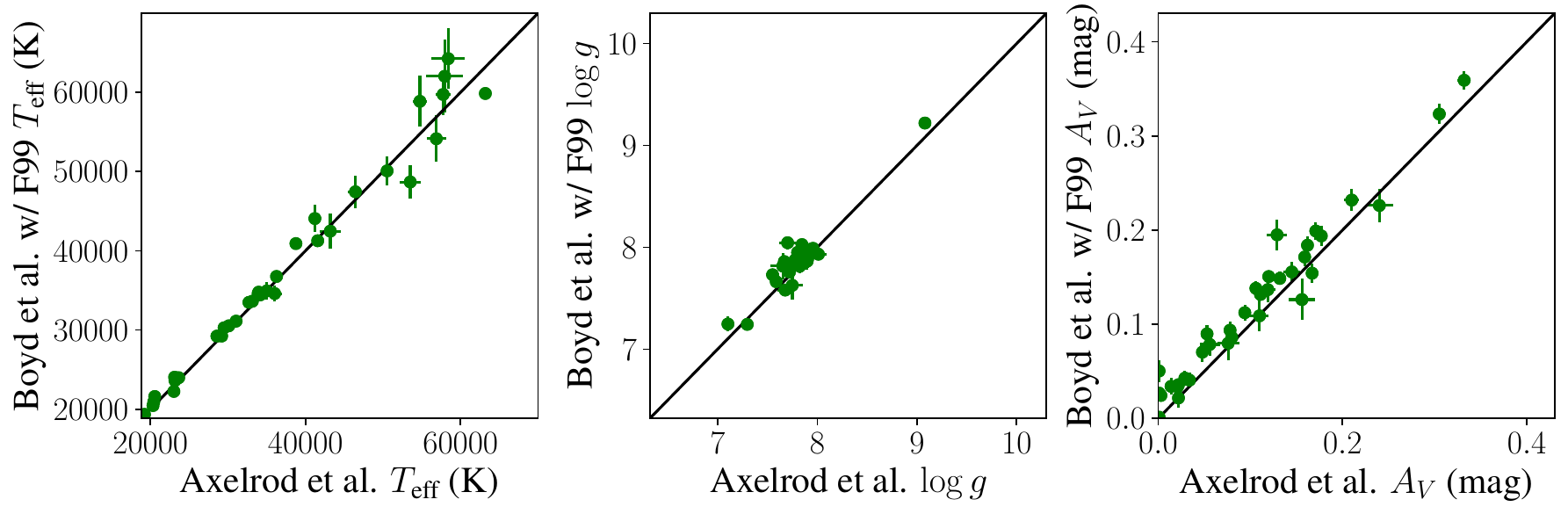}
    \caption{Comparing parameter inferences when using the same modelling decisions as \citet{a23}. This included using optical spectra only, using the \citet{Fitzpatrick1999} dust relation and a constant $R_V=3.1$.}
    \label{fig:a23_param_comp_f99}
\end{figure*}

\begin{landscape}
\begin{figure*}
\centering
	\includegraphics[width=1.\columnwidth]{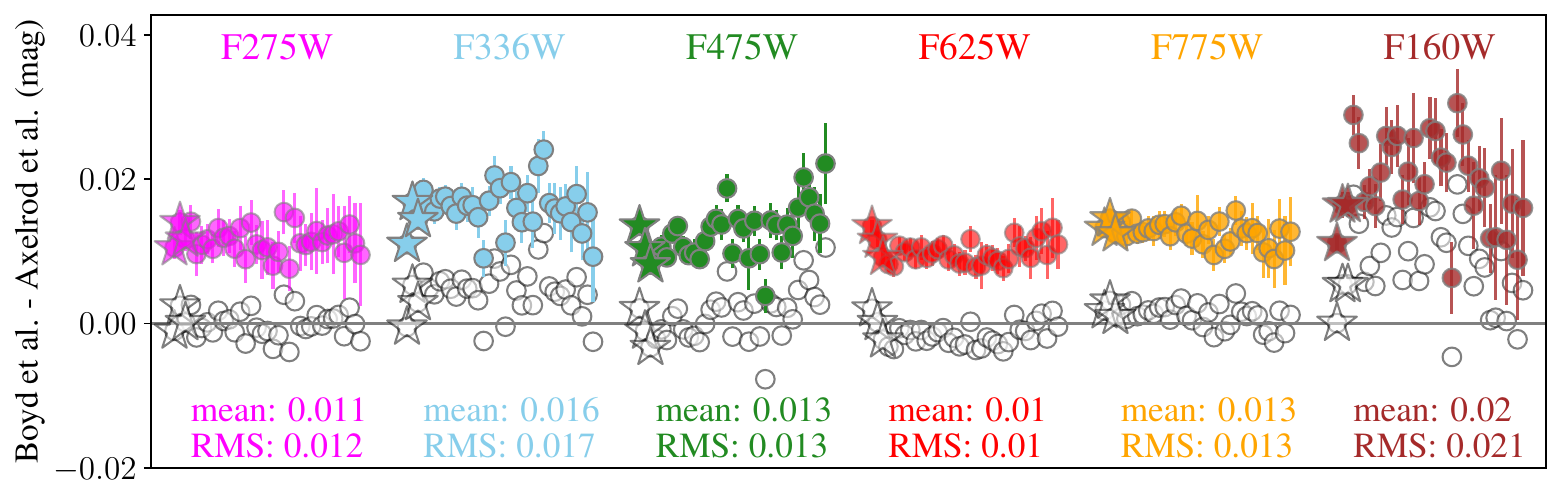}
    \caption{Residuals of synthetic HST photometry between our work and the previous work of \citet{a23} for the full network of 35 WDs. The work of \citet{a23} was tied to the older \citet{bohlin2014b} system. The coloured residuals and statistics are using our published synthetic photometry that is tied the updated \citet{bohlin2020} CALSPEC system. In white are the residuals when we use the \citet{bohlin2014b} CALSPEC system that matches what was used in \citet{a23}. Residuals shown are in AB mags and error bars represent the standard deviation of the inferred synthetic magnitudes from our work only. The unweighted mean and RMS scatter do not include the three primary standards that are illustrated by the star markers. }
    \label{fig:a23_phot_comp}
\end{figure*}
\end{landscape}

In Figure \ref{fig:a23_param_comp_f99} we compare parameter inferences between our model and \cite{a23} when we use the same model assumptions. We see that when we adopt the \cite{Fitzpatrick1999} dust relation, set a constant $R_V=3.1$ and ignore the HST/STIS UV spectra, the parameter inferences between the two models are consistent. This further supports the notion that differences in inferred SEDs are due to the differing model assumptions.

\newpage

\newpage

\section{White Dwarf Parameter Correlations}
\label{sed:wd_corr}
\begin{figure}
\centering
\includegraphics[width=0.8\columnwidth]{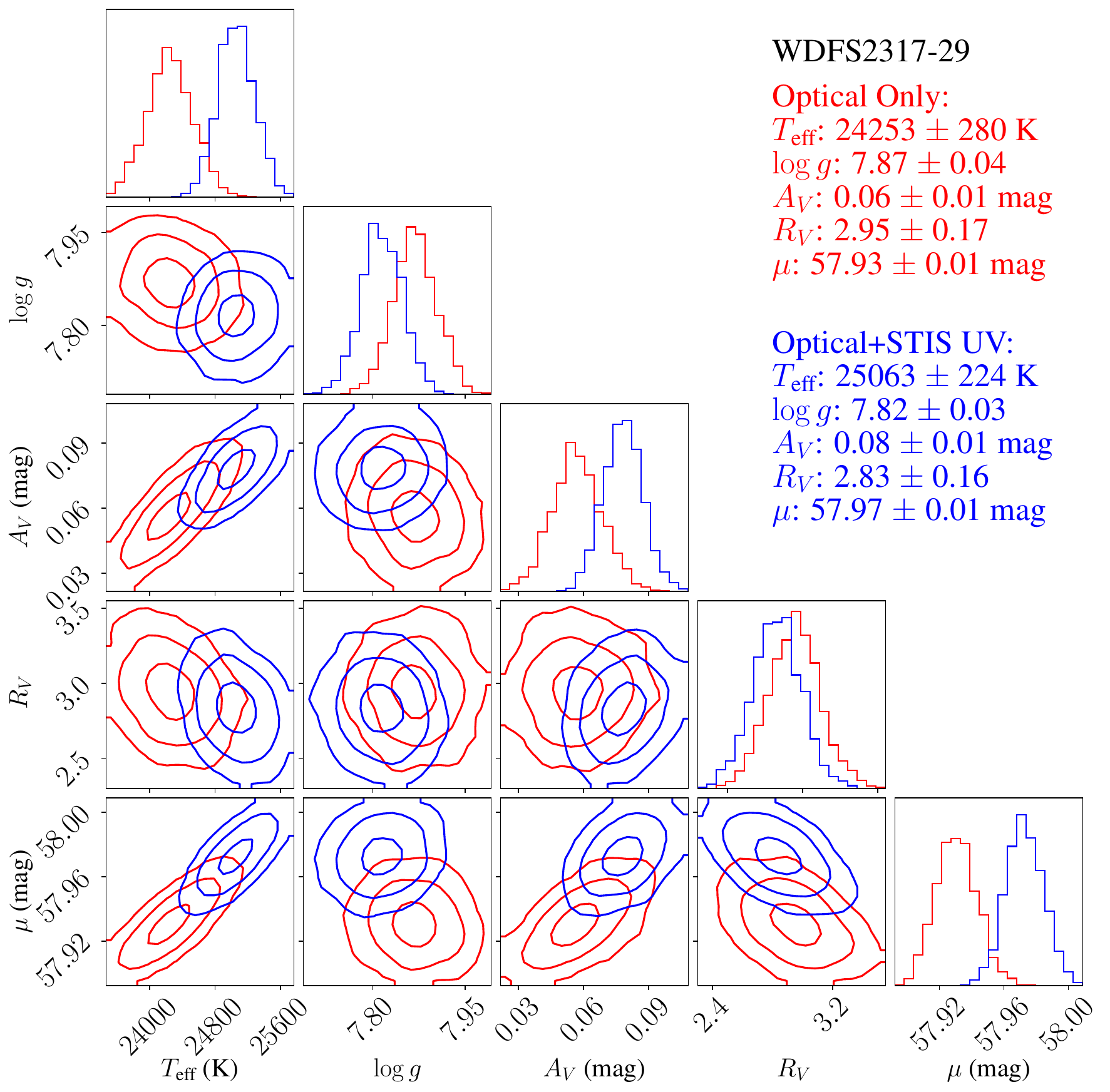}
    \caption{Corner plot showing the inferred parameter posteriors for WDFS2317-29. Posteriors in red are inferred using only the optical spectra, while the blue posteriors use both optical spectra and STIS UV spectra. In the posteriors we see strong correlations between the effective temperature $T_{\text{eff}}$, the dust extinction parameter $A_V$ and the achromatic offset parameter $\mu$.}
    \label{fig:wd_corner}
\end{figure}

In Figure \ref{fig:wd_corner}, we present the posterior distributions of the object-level parameters for WDFS2317-29. The posterior shows strong positive correlations among the effective temperature $T_{\text{eff}}$, the dust extinction parameter $A_V$ and the achromatic offset parameter $\mu$. Higher extinction diminishes absorption features, leading the model to infer higher effective temperatures as compensation to match the observed data. This trade-off between effective temperature and extinction preserves the colour of the WD but alters its observed brightness, which is then subsequently adjusted through the positive correlation with the achromatic offset parameter $\mu$. \cite{n19} used a similar methodology and found the same three-dimensional posterior correlations in Figure 14 of their work. The Bayesian analysis in \cite{ocallaghan2024} further explores trade-offs and degeneracies between dust and intrinsic stellar parameters using astrophysical priors and spectroscopic constraints.

Including the UV spectrum in the analysis raises the inferred effective temperature for WDFS2317-29, likely due to the additional information provided by the Lyman-$\alpha$ feature, which in turn increases the inferred extinction. As a result of the increased extinction and effective temperature, the inferred achromatic offset $\mu$ then rises to prevent an increase in the overall brightness of the synthetic photometry. The inclusion of STIS UV spectra, however, does not significantly affect the inference of $R_V$ for this object. Detecting differences in $R_V$ would require additional NIR spectra to capture the full panchromatic effects of dust.
\newpage

\section{HST/WFC3 Systematic Correlations}
\label{sec:sys_corr}
\begin{figure}
\centering
\includegraphics[width=0.9\columnwidth]{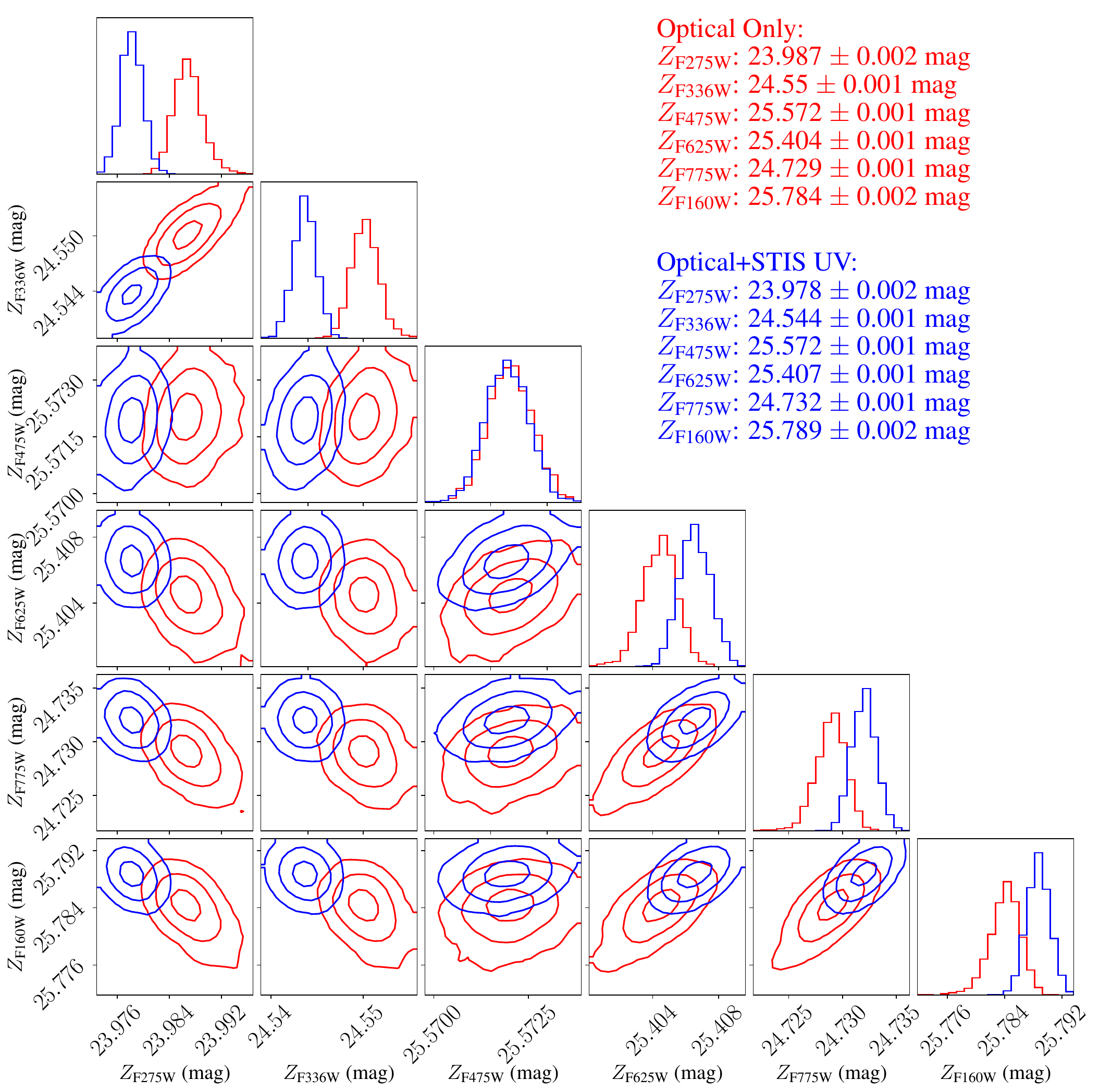}
    \caption{HST/WFC3 Cycle 25 UVIS1 and IR zeropoint inferences using only optical spectra in red and STIS UV spectra (as well as optical) in blue.}
    \label{fig:sys_corner}
\end{figure}

In Figure \ref{fig:sys_corner}, we plot the inferred posteriors for the HST/WFC3 photometric zeropoints. Including the STIS UV spectra has no effect on the F475W band, because this is where we set the CALSPEC primary magnitudes to constant. Adding the STIS UV spectra encourages the F275W and F336W zeropoints to be lower and the F675W, F775W, and F160W zeropoints to be higher. The changes in zeropoints are a reflection of the average effect that including the STIS spectra has on the inferred SEDs, as illustrated in Figure \ref{fig:stis_sed_comp}.

\begin{figure}
\centering
\hspace{1.7cm}
\includegraphics[width=0.55\columnwidth]{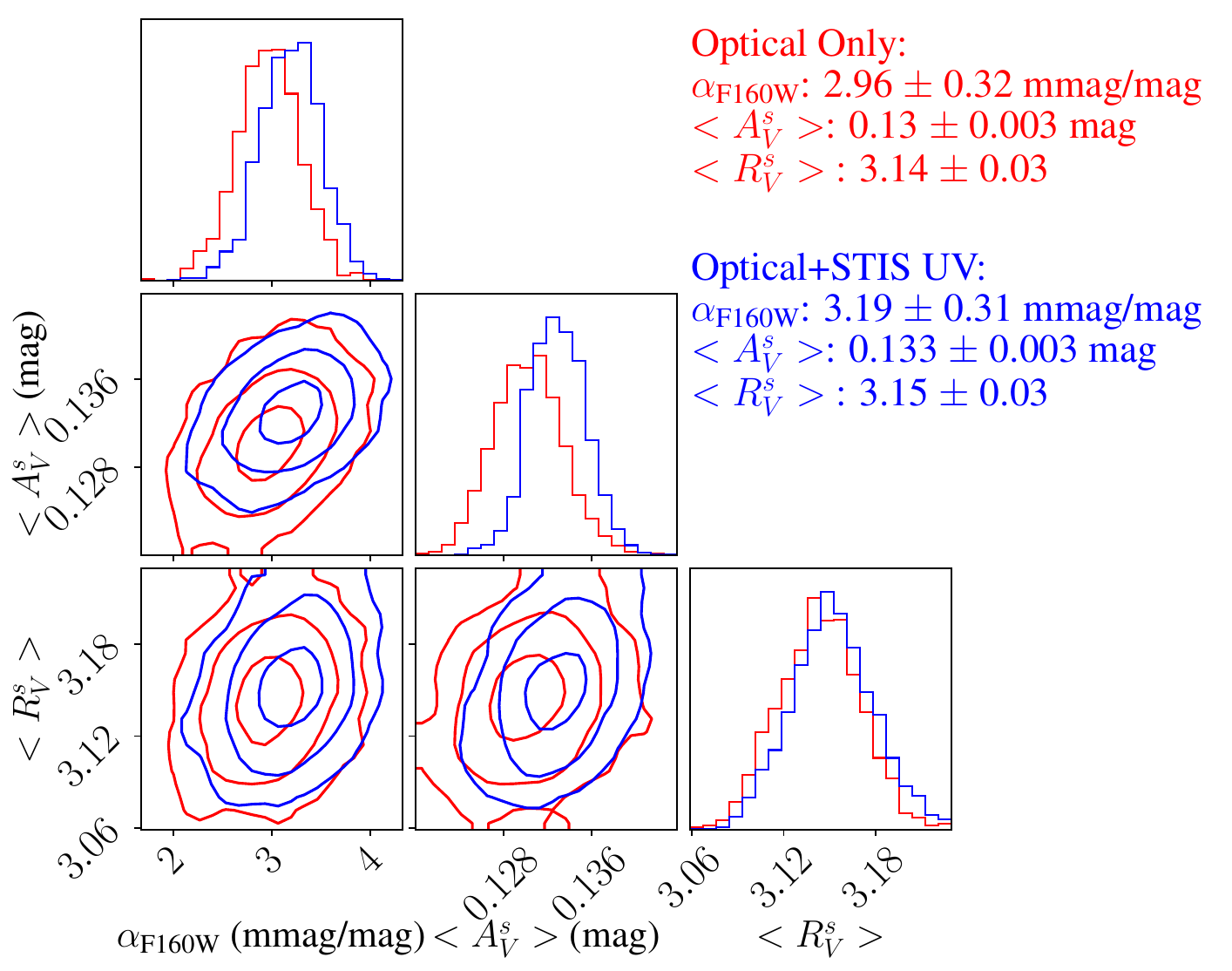}
    \caption{Corner plot showing the average inferred $A_V^s$ and $R_V^s$ over 2000 posteriors samples along with the inferred HST/WFC3 F160W count-rate nonlinearity (CRNL). The plot shows a positive correlation between the average inferred $A_V^s$ and the inferred CRNL.}
    \label{fig:sys_corner2}
\end{figure}

We examine the correlation between dust and the inferred F160W CRNL in Figure \ref{fig:sys_corner2}. A positive correlation is observed between the average of inferred $A_V^s$ values and inferred CRNL value. This suggests that the CRNL inferred using this method is sensitive to the chosen dust relation and any additional assumptions applied to it. Similar sensitivity to the dust relation was previously noted in Section \ref{sec:hst_sys} when inferring a CRNL closer to the result reported in \cite{a23} when using the \cite{Fitzpatrick1999} dust relation.

\newpage
\section{Population Dust Constraints}

\begin{figure}
\centering
\includegraphics[width=0.45\columnwidth]{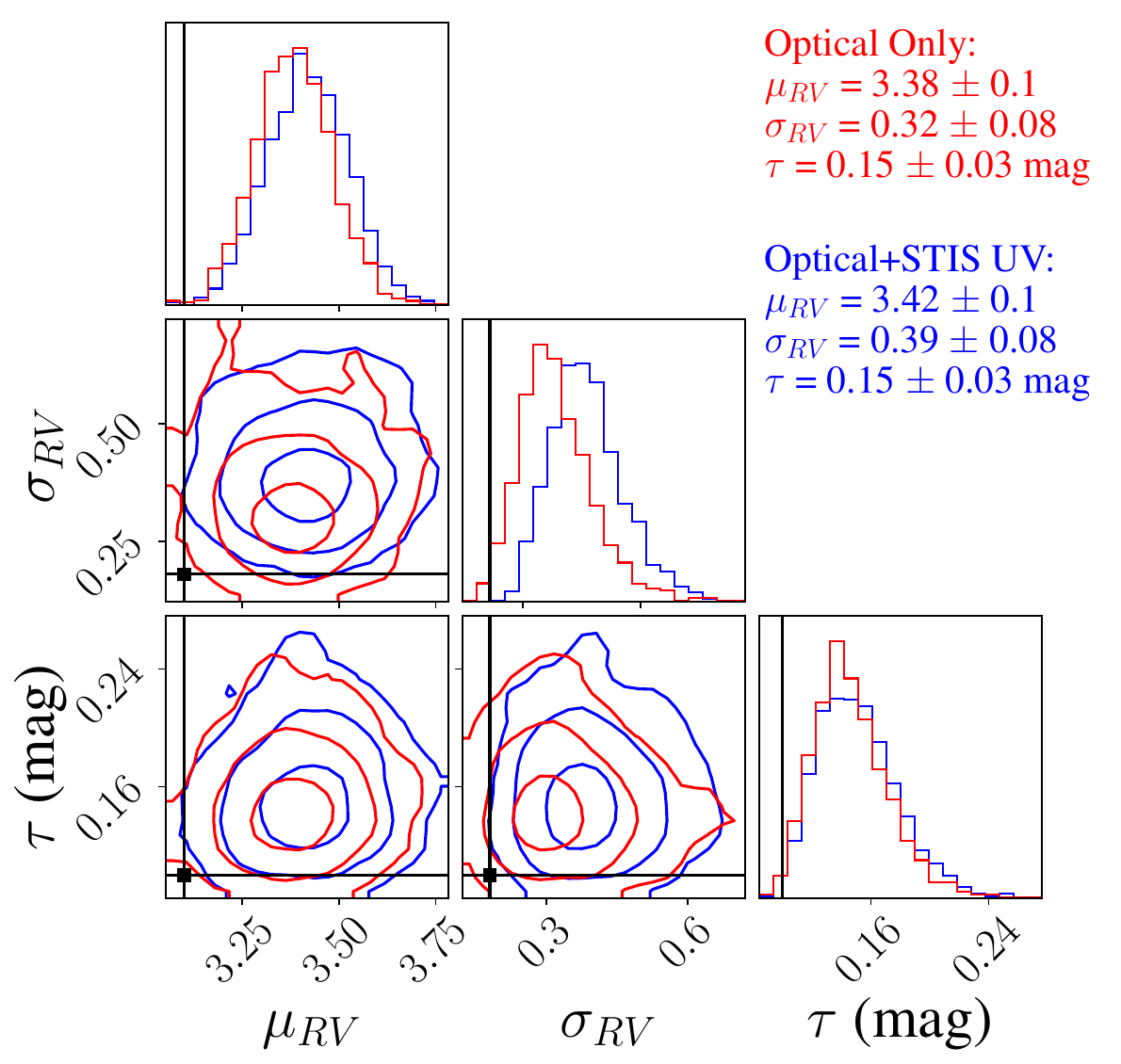}
    \caption{Population dust parameters inferred with (blue) and without (red) the STIS UV spectra for the 19 WDs that have it. These results are only a demonstration of the hierarchical capabilities of the model and are influenced by selection effects. We show in black the constant population values motivated by literature that we use in the priors for our main analysis.}
    \label{fig:rv_corner}
\end{figure}
As a demonstration of our models ability to infer population-level parameters, we infer a $\mu_{RV}$ and $\sigma_{RV}$ given a truncated normal prior $R^s_V\sim \mathcal{TN}(\mu_{RV},\sigma_{RV},1.2,\infty)$. We also infer an average population extinction $\tau$ assuming $A^s_V$ follows an exponential distribution. The resulting posteriors with and without using the HST/STIS UV spectra are shown in Figure \ref{fig:rv_corner}. The inferred parameters are significantly different when compared to the constant population parameters used in our main analysis motivated by the work of \cite{schlafly2016}. The results in our population inference are to be taken with caution, as our DA WD network is strongly influenced by selection effects that are not taken into account in the analysis. In Section \ref{sec:future} we discuss future plans to apply the model to a larger volume limited sample of DA WDs to provide constraints on these population parameters that are uninfluenced by selection effects.
The results show that the introduction of the HST/STIS UV spectra does not significantly change the $R_V$ population parameters. We would only expect substantial difference if we obtain NIR spectra for a panchromatic view to monitor the effects of the dust relations. 
\newpage

\section{HST Zeropoint Shifts}
\label{sec:zp_shift}
\begin{figure}
\centering
\includegraphics[width=0.9\columnwidth]{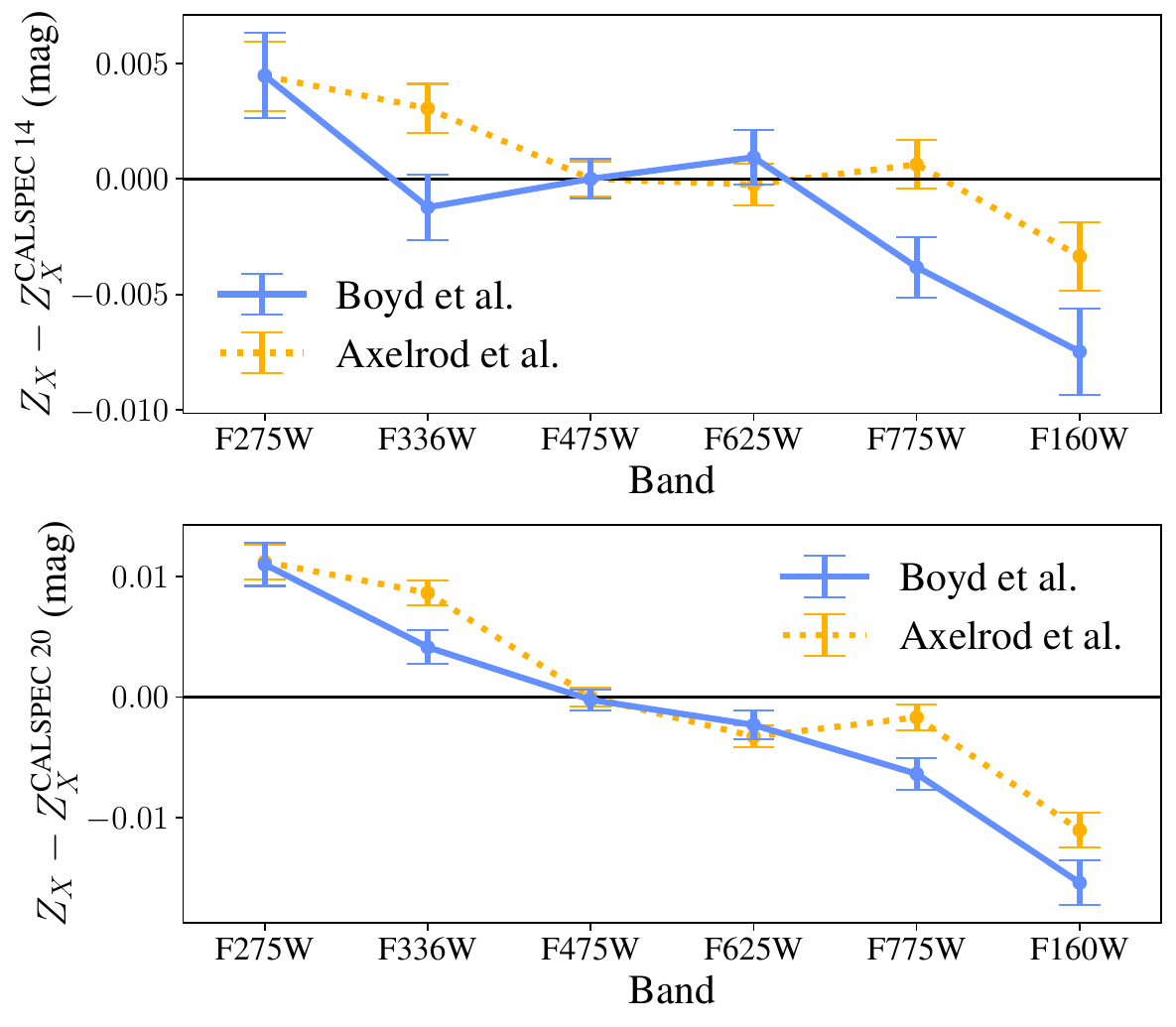}
    \caption{Relative HST/WFC3 zeropoint deviations away from the CALSPEC system for this work (presented as the solid blue line) and \citet{a23} (presented as the orange dotted line). As a post-hoc step we arbitrarily to tie our relative zeropoints to the CALSPEC system at the F475W band using the difference between observed a synthetic magnitudes from \citet{bohlin2014b} (top plot) and \citet{bohlin2020} (bottom plot) for the three primary standards. The zeropoints calculated are AB magnitudes for UVIS1 and IR detectors during Cycle 25 (average MJD=57588.79 days) only.}
    \label{fig:zp_shift}
\end{figure}

A key advantage of the hierarchical aspect of our work and the work of \cite{a23} is that we can infer relative HST/WFC3 offsets that deviate from the absolute CALSPEC system. To derive the CALSPEC zeropoints for Cycle 25 (average MJD=57588.79 days) we adopt the same scheme as \cite{n19} where we use the difference between synthetic magnitudes and what is observed for the three primary stars only. In Table \ref{tab:zp_shift} we present zeropoints derived post-hoc by tying our relative zeropoints and those derived in \cite{a23} to the F475W  absolute zeropoint in the \cite{bohlin2014b} and \cite{bohlin2020} CALSPEC systems. Using the achromatic differences between the zeropoints in the table we can easily convert our network between systems without needing to reanalyse our data. In Figure \ref{fig:zp_shift} we plot the deviations from the two CALSPEC systems. We see that our work is in good agreement with results inferred by \cite{a23} in the UV, with relative zeropoints deviating from CALSPEC zeropoints by around 11 mmag in the 2020 system and 4.5 mmag in the 2014 system. In the NIR we infer larger relative zeropoints shifts with CALSPEC with differences up to 7.5 mmag for the 2014 system and 15 mmag for the 2020 system. Despite an independent analysis, methodology, and dataset, our relative zeropoints show the same general trend as \cite{a23}, decreasing as a function of wavelength. We expect our relative zeropoints to differ of this order from CALSPEC since we have analysed a larger and fainter set of standards.

\begin{landscape}
\begin{table*}
	\centering	
    \caption{In this table we present inferred  HST/WFC3 zeropoints for this work and \citet{a23} in comparison to two CALSPEC systems.  }
\label{tab:zp_shift}
	\begin{tabular}{lcccccc} %
\hline
\multicolumn{7}{c}{CALSPEC Bohlin (2014)}\\
\hline
&F275W&F336W&F475W&F625W&F775W&F160W\\
\hline
Boyd et al. $Z_X$ (mag)&23.9899 $\pm$ 0.0017&24.5553 $\pm$ 0.0012&25.5836 $\pm$ 0.0006&25.4182 $\pm$ 0.0009&24.7436 $\pm$ 0.0011&25.8004 $\pm$ 0.0016\\
Axelrod et al. $Z_X$ (mag)&23.9898 $\pm$ 0.0013&24.5596 $\pm$ 0.0008&25.5836 $\pm$ 0.0004&25.417 $\pm$ 0.0006&24.7481 $\pm$ 0.0008&25.8045 $\pm$ 0.0011\\
\hline
 $Z_X^{\text{CALSPEC 14}}$ (mag)&23.9854 $\pm$ 0.0008&24.5565 $\pm$ 0.0007&25.5836 $\pm$ 0.0006&25.4173 $\pm$ 0.0007&24.7475 $\pm$ 0.0007&25.8078 $\pm$ 0.001\\
\hline
Boyd et al. $Z_X-Z_X^{\text{CALSPEC 14}}$ (mmag)&4.5 $\pm$ 1.8&-1.2 $\pm$ 1.4&0.0 $\pm$ 0.9&0.9 $\pm$ 1.2&-3.8 $\pm$ 1.3&-7.5 $\pm$ 1.9\\
Axelrod et al. $Z_X-Z_X^{\text{CALSPEC 14}}$ (mmag)&4.4 $\pm$ 1.5&3.1 $\pm$ 1.1&0.0 $\pm$ 0.8&-0.3 $\pm$ 0.9&0.6 $\pm$ 1.1&-3.4 $\pm$ 1.5\\
\hline
\multicolumn{7}{c}{CALSPEC Bohlin et al. (2020)}\\
\hline
&F275W&F336W&F475W&F625W&F775W&F160W\\
\hline
Boyd et al. $Z_X$ (mag)&23.9782 $\pm$ 0.0017&24.5437 $\pm$ 0.0012&25.5719 $\pm$ 0.0006&25.4066 $\pm$ 0.0009&24.732 $\pm$ 0.0011&25.7887 $\pm$ 0.0016\\
Axelrod et al. $Z_X$ (mag)&23.9784 $\pm$ 0.0013&24.5482 $\pm$ 0.0008&25.5722 $\pm$ 0.0004&25.4056 $\pm$ 0.0006&24.7367 $\pm$ 0.0008&25.7931 $\pm$ 0.0011\\
\hline
 $Z_X^{\text{CALSPEC 20}}$ (mag)&23.9672 $\pm$ 0.0006&24.5395 $\pm$ 0.0007&25.5722 $\pm$ 0.0006&25.4089 $\pm$ 0.0007&24.7384 $\pm$ 0.0007&25.8041 $\pm$ 0.0009\\
\hline
Boyd et al. $Z_X-Z_X^{\text{CALSPEC 20}}$ (mmag)&11.0 $\pm$ 1.8&4.1 $\pm$ 1.4&-0.2 $\pm$ 0.9&-2.3 $\pm$ 1.2&-6.3 $\pm$ 1.3&-15.4 $\pm$ 1.8\\
Axelrod et al. $Z_X-Z_X^{\text{CALSPEC 20}}$ (mmag)&11.2 $\pm$ 1.4&8.6 $\pm$ 1.0&0.0 $\pm$ 0.7&-3.3 $\pm$ 0.9&-1.7 $\pm$ 1.1&-11.0 $\pm$ 1.4\\
\hline
\end{tabular}
\begin{flushleft}\small
\textbf{Notes:}  We calculate two sets of zeropoints for both the \cite{bohlin2014b} and the \cite{bohlin2020} CALSPEC systems. CALSPEC zeropoints and uncertainties are inferred using the same scheme as \cite{n19} where we use the difference between synthetic magnitudes (derived from published SEDs) and what we observe in Cycle 25 for the three primary standards only. Our main results are already with respect to the CALSPEC 2020 system, tied to the F475W band. To move our results onto the CALSPEC 2014 system we simply add the difference between 2020 and 2014 systems in the F475W band to all bands. To place the \cite{a23} relative zeropoints onto the CALSPEC 2014 system, we subtract their inferred changes in zeropoints
(shifted so that the F475W is at zero) to the derived CALSPEC 2014 zeropoints. To then place \citet{a23} zeropoints onto the CALSPEC 2020 system, we subtract the difference between 2020 and 2014 systems in the F475W band from all the bands. Presented zeropoints are equivalent to the UVIS1 and IR zeropoints during Cycle 25 (average MJD=57588.79 days) in AB magnitudes. Zeropoints are also calculated from aperture radii (7.5 pixels for UVIS and 5 pixels for the IR) to infinity.
\end{flushleft}
\end{table*}
\end{landscape}
\newpage
\section{Inferred Spectroscopic Spline Residuals}
\begin{figure}
\centering
\includegraphics[width=1\columnwidth]{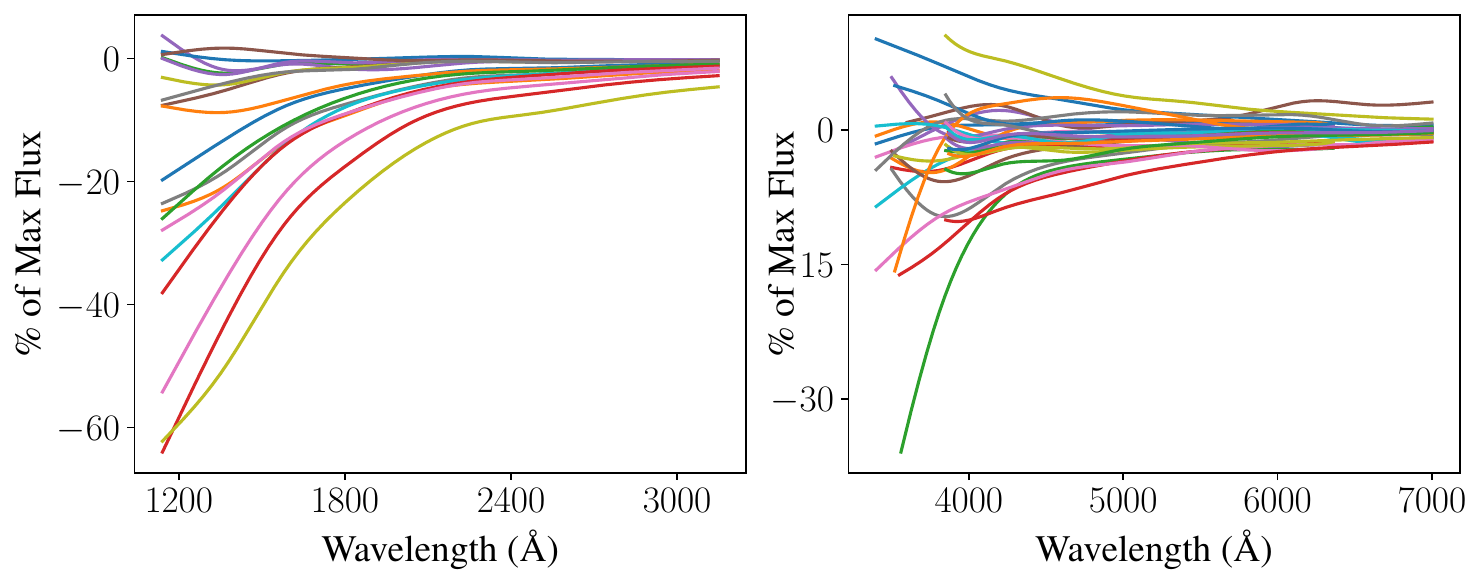}
    \caption{Inferred spline residuals for the STIS UV spectra (left) and the ground-based optical spectra (right). Spline strength is expressed as the percentage of the maximum flux value in its respective spectrum. There is no meaning behind the colour of each spline.}
    \label{fig:splines}
\end{figure}

To analyse the spectroscopic data of our network we add a cubic spline to the inferred SED in flux space. In Figure \ref{fig:splines} we over-plot the inferred splines for both the UV STIS spectra and the optical spectra. We interpret the optical functions as reflecting instrumental systematics, with flux corrections being more biased near the spectrograph edges where throughput declines sharply. For the STIS UV spectra we see strong spline functions, particularly at shorter wavelengths. These results may be indicative of STIS systematics, resulting in an underestimation of flux. Alternatively, this may be understood as a difference between model and data, where perhaps our chosen dust relation needs improvement or our theoretical DA WD template needs refining. In Section \ref{sec:splines} we discuss various STIS systematics and caveats in dust relations that may have led to these inferred splines.

\newpage
\begin{landscape}
\section{Residuals with different modelling assumptions}
\begin{table*}
	\centering	
    \caption{HST/WFC3 residual mean $\mu$ and RMS for the 32 faint DA WDs when comparing their observed photometry with synthetic photometry for different experiments. Opt only signifies that we use only ground-based optical spectra and HST photometry to calibrate the stars. Opt+STIS signifies that we also use the STIS UV spectra for the 19 WDs that had coverage. G23 indicates that we use the \citet{gordon2023} dust relation, whilst F99 represents the \citet{Fitzpatrick1999} dust relation. Const $R_{V}$ refers to a constant $R_{V}=3.1$. Norm $R_{V}$ experiments used the truncated normal $R_{V}$ prior $\mathcal{TN}(\mu_{RV}=3.1,\sigma_{RV}=0.18,1.2,\infty)$. Population $R_{V}$ experiments also inferred the population parameters $\mu_{RV}$, $\sigma_{RV}$ and $\tau$. We label tests presented in our final results with an $*$. Results shown are in AB millimagnitudes.}
\label{tab:tests}
	\begin{tabular}{lcccccccccccccc} 
 
		\hline
		Test  
        &\multicolumn{2}{c}{All}&\multicolumn{2}{c}{F275W}&\multicolumn{2}{c}{F336W}&\multicolumn{2}{c}{F475W}&\multicolumn{2}{c}{F625W}&\multicolumn{2}{c}{F775W}&\multicolumn{2}{c}{F160W}  \\
        & $\mu$ & RMS &$\mu$ & RMS &$\mu$ & RMS &$\mu$ & RMS &$\mu$ & RMS &$\mu$ & RMS  
        &$\mu$ & RMS \\
        \hline

Opt Only G23 Const $R_V$&-0.01&4.36&-0.13&3.54&-0.11&3.44&1.52&4.34&-0.36&2.35&0.1&3.48&-1.09&7.3\\
Opt Only G23 Norm $R_V$&-0.04&3.79&0.02&2.25&-0.28&2.77&1.43&4.09&-0.37&2.3&0.22&3.44&-1.29&6.29\\
Opt Only G23 Pop $R_V$&0.02&3.67&0.65&2.33&-0.81&2.65&1.12&3.74&-0.39&2.35&0.85&4.37&-1.29&5.47\\
Opt+STIS G23 Const $R_V$&-0.02&4.68&0.16&3.47&0.26&3.56&1.35&4.23&-0.64&2.4&0.03&3.44&-1.31&8.43\\
Opt+STIS G23 Norm $R_V$$^{*}$&-0.06&3.9&0.38&2.44&-0.07&3.29&1.2&3.78&-0.62&2.33&0.18&3.43&-1.44&6.56\\
Opt+STIS G23 Pop $R_V$&-0.07&3.6&1.16&3.03&-0.57&3.27&0.82&3.27&-0.67&2.4&0.42&3.38&-1.55&5.48\\
\hline
Opt Only F99 Const $R_V$&0.05&4.59&1.42&4.27&-1.29&3.79&0.7&3.77&-0.21&2.32&0.56&3.51&-0.89&7.86\\
Opt Only F99 Norm $R_V$&0.01&4.07&1.26&3.33&-1.17&3.13&0.64&3.77&-0.2&2.24&0.6&3.45&-1.09&6.9\\
Opt Only F99 Pop $R_V$&-0.01&3.85&1.3&3.39&-1.14&2.85&0.7&3.71&-0.18&2.2&0.5&3.33&-1.26&6.31\\
Opt+STIS F99 Const $R_V$&-0.06&4.81&1.38&4.35&0.04&3.8&-0.73&3.65&-0.67&2.5&1.23&3.61&-1.62&8.53\\
Opt+STIS F99 Norm $R_V$&-0.08&4.09&1.05&3.13&0.13&3.43&-0.66&3.5&-0.6&2.31&1.09&3.57&-1.47&6.96\\
Opt+STIS F99 Pop $R_V$&-0.08&3.84&0.84&3.19&0.33&3.08&-0.31&3.14&-0.43&2.25&0.7&3.24&-1.61&6.57\\

  \hline

\end{tabular}
\end{table*}

In our work we make various modelling assumptions that affect our ability to reproduce the data. An example of such a decision is whether we only use optical spectra in our inference, or whether we also use the STIS UV spectra for the 19 WDs that have coverage. Further assumptions were related to dust where we use the \cite{gordon2023} relation, whereas previous analyses have used \cite{Fitzpatrick1999}. Finally, our priors on $R_V$ are important to consider, in previous analysis this has been kept constant as $R_V=3.1$, but in our work we use a truncated normal prior and also experiment with using a population prior. In Table \ref{tab:tests} we show how varying these assumptions affects our residual bias and scatter with HST/WFC3 photometry.

We find that RMS scatter is lower overall when we use the updated \cite{gordon2023} relation over the \cite{Fitzpatrick1999} relation. Including the STIS UV spectra makes the RMS scatter marginally worse in all bands except the F475W. We also find that granting the model additional flexibility to infer population parameters for the $R_V$ prior significantly reduces the RMS scatter. Our goal is not solely to minimise RMS residuals with the HST/WFC3 photometry; rather, we aim to leverage all available data while adhering to physically motivated modelling principles. Accordingly, we present our final results based on analyses that incorporates both the optical and STIS UV spectra, utilises the updated \cite{gordon2023} dust relation, and adopts a truncated normal $R_V$ prior informed by the work of \cite{schlafly2016}.

\end{landscape}

\newpage
\section{Photometric Residuals with different surveys}

\begin{figure*}
\centering
\includegraphics[width=1.\columnwidth]{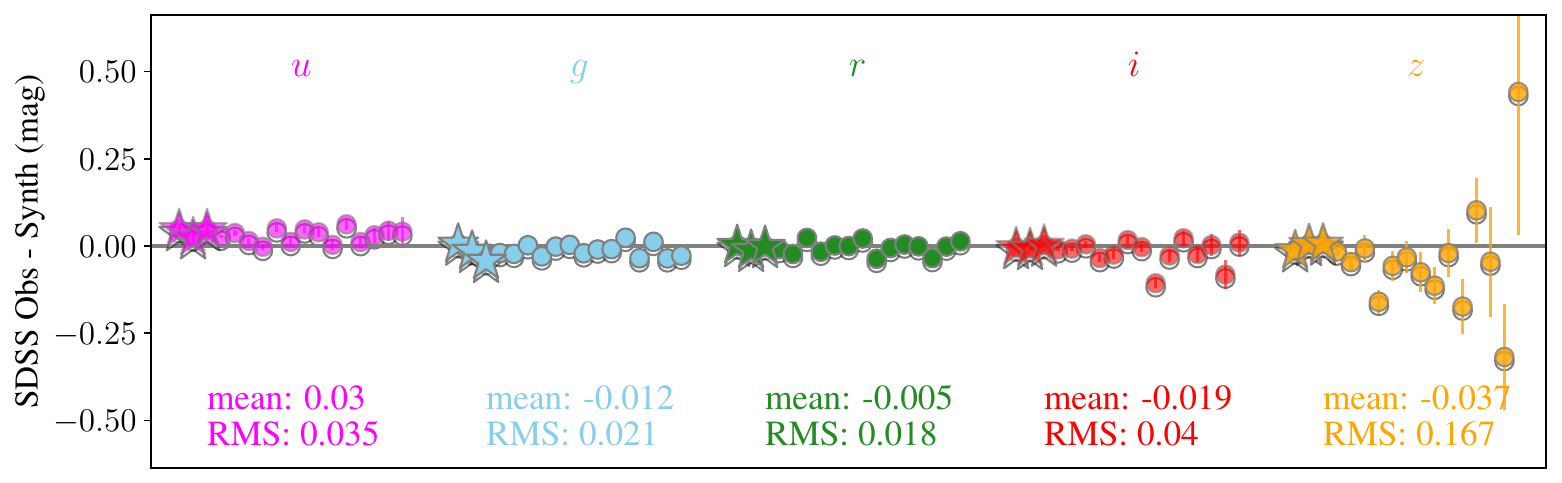}
    \caption{Photometric residuals when comparing our synthetic SDSS photometry with real SDSS observations \citep{holberg2006} for the faint DA WD standards that had coverage. Objects are arranged in order of brightness, ending with the dimmest object. We presented the unweighted mean and RMS residual for each band. Residuals are in AB magnitudes. Plotted underneath in white are the residuals when using the \citet{bohlin2014b} CALSPEC system rather than the \citet{bohlin2020} system.}
    \label{fig:sdss_resid}
\end{figure*}

\begin{figure*}
\centering
	\includegraphics[width=1.\columnwidth]{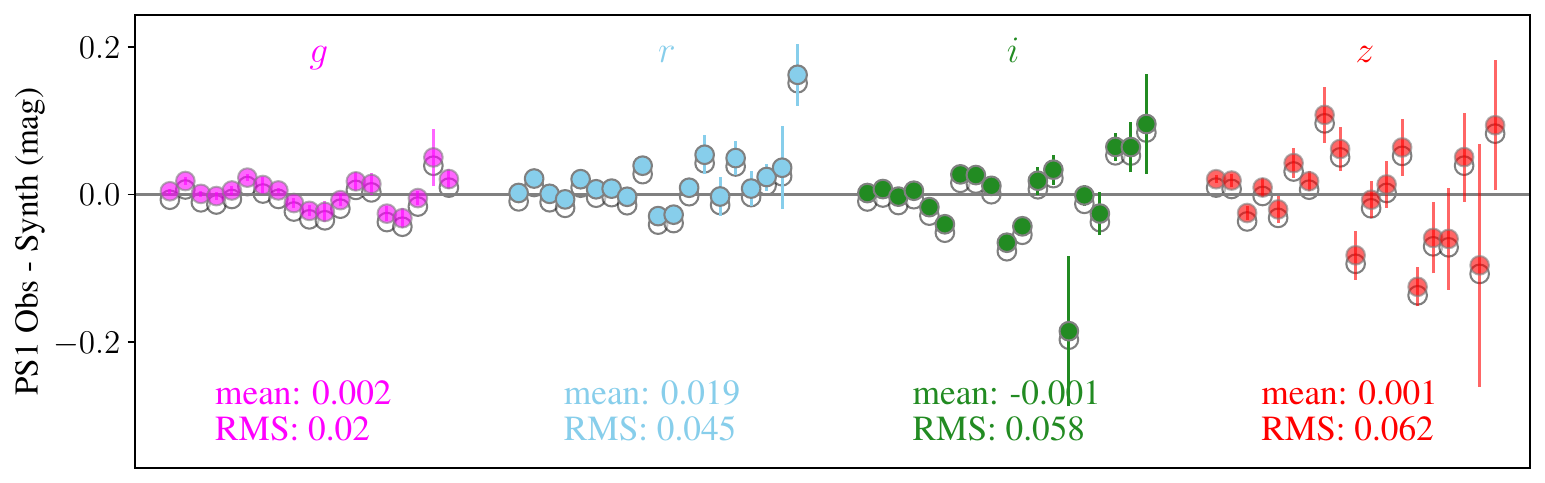}
    \caption{Photometric residuals when comparing our synthetic PS1 photometry with real PS1 observations \citep{ps1_data} for the faint DA WD standards that had coverage. Objects are arranged in order of brightness, ending with the dimmest object. We presented the unweighted mean and RMS residual for each band. Residuals are in AB magnitudes. Plotted underneath in white are the residuals when using the \citet{bohlin2014b} CALSPEC system rather than the \citet{bohlin2020} system.}
    \label{fig:ps1_resid}
\end{figure*}

\begin{figure*}
\centering
	\includegraphics[width=1.\columnwidth]{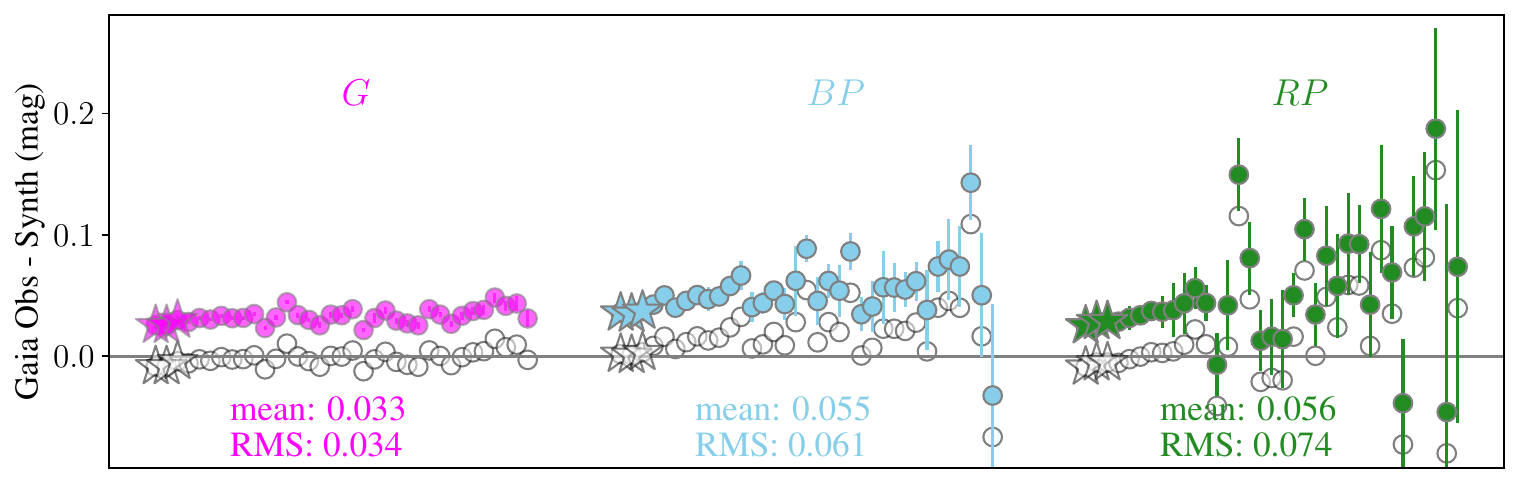}
    \caption{Photometric residuals when comparing our synthetic \textit{Gaia} photometry with real \textit{Gaia} observations \citep{gaia_dr3} for the faint DA WD standards that had coverage. Objects are arranged in order of brightness, ending with the dimmest object. We presented the unweighted mean and RMS residual for each band. Residuals are in Vega magnitudes. Plotted in white are the residuals when using the \citet{bohlin2014b} CALSPEC system (rather than the \citet{bohlin2020} system) and \textit{Gaia}'s scaled definition of Vega.}
    \label{fig:gia_resid}
\end{figure*}

\begin{figure*}
\centering
	\includegraphics[width=1.\columnwidth]{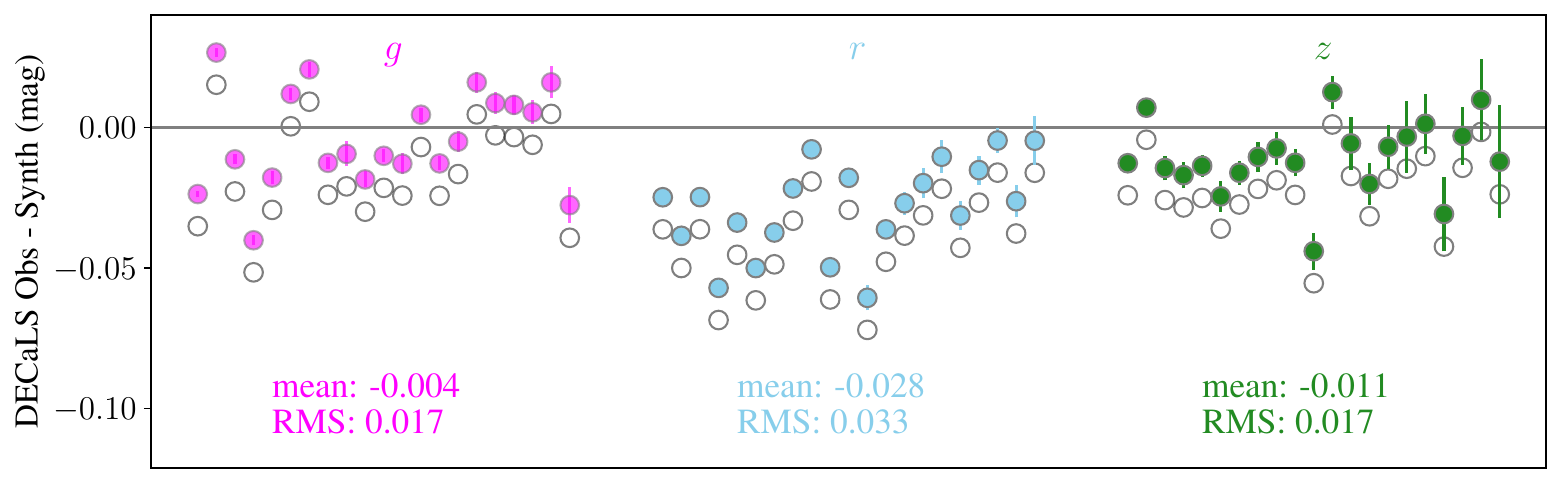}
    \caption{Photometric residuals when comparing our synthetic DECaLS photometry with real DECaLS observations \citep{decals} for the faint DA WD standards that had coverage. Objects are arranged in order of brightness, ending with the dimmest object. We presented the unweighted mean and RMS residual for each band. Residuals are in AB magnitudes. Plotted in white are the residuals when using the \citet{bohlin2014b} CALSPEC system rather than the \citet{bohlin2020} system.}
    \label{fig:decals_resid}
\end{figure*}

Using our inferred synthetic SEDs for the WDs, we can integrate over various bandpass filters corresponding to different surveys. For any survey that observed a given standard, we can compute the residual between their measurement and our synthetic prediction. We show our residuals for DES in Figure \ref{fig:des_resid}, SDSS in Figure \ref{fig:sdss_resid}, PS1 in Figure \ref{fig:ps1_resid}, \textit{Gaia} in Figure \ref{fig:gia_resid} and DECaLS in Figure \ref{fig:decals_resid}. The data collection process and residual results are discussed in Section \ref{sec:other_res}. Since each survey is calibrated independently, a wavelength-dependent bias is expected. Additional biases are anticipated because our SEDs are tied to the new CALSPEC \cite{bohlin2020} system rather than \cite{bohlin2014b}. In the Figures we show that the residuals are significantly smaller, comparable to those in \cite{a23}, when we tie our network to the the older CALSPEC system \citep{bohlin2014b}. In Figure \ref{fig:gia_resid} we also find smaller residuals when using the same scaled definition of Vega as \textit{Gaia} \citep{riello2021}. Despite getting better agreement with other surveys when using the older CALSPEC system, we maintain our decision to publish the network with respect to the more accurate \cite{bohlin2020} CALSPEC system and encourage other surveys to adopt the same for their absolute calibration. Quantifying these biases is crucial for cross-calibrating surveys and combining samples. In Section \ref{sec:future}, we discuss the ongoing application of our SEDs to unify SN Ia samples across different surveys. Synthetic photometry and residuals for these surveys can be found at: \url{https://zenodo.org/records/16877348/files/paper_tables.zip}.

\newpage




\bsp	
\label{lastpage}
\end{document}